\documentclass[aps,prd,showpacs,superscriptaddress,nofootinbib,groupedaddress]{revtex4}
\pdfoutput=1

\usepackage{pstricks}
\usepackage{color}
\usepackage{graphicx}
\usepackage{amsmath} 
\usepackage[colorlinks=false, urlcolor=blue, pdfborder={0 0 0}]{hyperref}
\usepackage{amssymb}
\usepackage{ulem}

\graphicspath{{figs/}}

\def\gsim{\raise0.3ex\hbox{$\;>$\kern-0.75em\raise-1.1ex\hbox{$\sim\;$}}}
\def\lsim{\raise0.3ex\hbox{$\;<$\kern-0.75em\raise-1.1ex\hbox{$\sim\;$}}}
\def\beqn#1{\begin{equation}\label{#1}}
\def\eeqn{\end{equation}}
\def\beqa#1{\begin{eqnarray}\label{#1}}
\def\eeqa{\end{eqnarray}}

\def\SM{$\mathrm{SU(3)_c \otimes SU(2)_L \otimes U(1)_Y}$ }
\def\21{$\mathrm{SU(2)_L \otimes U(1)_Y}$ }

\def\lnv{lepton number violation }

\def\vev#1{\left\langle #1\right\rangle}
 \usepackage{color}
 \usepackage{enumerate}
\definecolor{nicered}{rgb}{0.7,0.1,0.1}
\definecolor{nicegreen}{rgb}{0.1,0.5,0.1}
\hypersetup{colorlinks,citecolor= nicegreen,linkcolor= nicered}
\nopagebreak
\allowdisplaybreaks

  \newcommand{\AddrLiege}{{\sl \small IFPA, Dep. AGO, Universite de
      Liege, Bat B5,\\ \small \sl Sart Tilman B-4000 Liege 1,
      Belgium}}
  \newcommand{\AddrAHEP}{%
  AHEP Group, Institut de F\'{i}sica Corpuscular --
  C.S.I.C./Universitat de Val\`{e}ncia, Parc Cientific de Paterna.\\
 C/ Catedratico Jose Beltran, 2 E-46980 Paterna (Val\`{e}ncia) - SPAIN}

\begin{document}
\title{Leptogenesis with a dynamical seesaw scale} 
\author{D. Aristizabal Sierra}
\email{daristizabal@ulg.ac.be}
\affiliation{\AddrLiege} 
\author{M. T\'ortola}
\email{mariam@ific.uv.es}
\affiliation{\AddrAHEP}
\author{J. W. F. Valle}
\email{valle@ific.uv.es}
\affiliation{\AddrAHEP}
\author{A. Vicente}
\email{Avelino.Vicente@ulg.ac.be}
\affiliation{\AddrLiege}
\date{\today}
\vspace*{0.2cm}
\begin{abstract}
  %
  In the simplest type-I seesaw leptogenesis scenario right-handed
  neutrino annihilation processes are absent. However, in the presence
  of new interactions these processes are possible and can affect the
  resulting $B-L$ asymmetry in an important way. A prominent example
  is provided by models with spontaneous lepton number violation,
  where the existence of new dynamical degrees of freedom can play a
  crucial role.
  In this context, we provide a model-independent discussion of the
  effects of right-handed neutrino annihilations.
  We show that in the weak washout regime, as long as the scattering
  processes remain slow compared with the Hubble expansion rate
  throughout the relevant temperature range, the efficiency can be
  largely enhanced, reaching in some cases maximal values. Moreover,
  the $B-L$ asymmetry yield turns out to be independent upon initial
  conditions, in contrast to the ``standard'' case. On the other hand,
  when the annihilation processes are fast, the right-handed neutrino
  distribution tends to a thermal one down to low temperatures,
  implying a drastic suppression of the efficiency which in some cases
  can render the $B-L$ generation mechanism inoperative.
  \end{abstract}
\maketitle
\setcounter{footnote}{0}

\section{Introduction}
\label{sec:intro}

The only indications for physics beyond the \21 model come from the
lepton sector and cosmology.  In particular, the need to account for
neutrino mass~\cite{Tortola:2012te} as well as the cosmological baryon
asymmetry~\cite{Hinshaw:2012aka,Ade:2013zuv} have brought substantial
interest on different variants of the seesaw mechanism with
high~\cite{minkowski:1977sc,gell-mann:1980vs,yanagida:1979,
  Lazarides:1980nt,mohapatra:1980ia,Schechter:1980gr} as well as low
\lnv
scale~\cite{mohapatra:1986bd,akhmedov:1995ip,akhmedov:1995vm,Malinsky:2005bi}.
Like the electroweak gauge symmetry, it is reasonable to imagine that
also lepton number symmetry is broken spontaneously in order to
generate neutrino masses.
A simple framework for this is the \21 seesaw scenario with
spontaneous violation of ungauged lepton
number~\cite{chikashige:1981ui,Schechter:1981cv,gelmini:1984ea,gonzalezgarcia:1989rw},
whose implementation requires the presence of a lepton-number-carrying
complex scalar singlet coupled to the electroweak singlet
right-handed (RH) neutrinos.
Its imaginary part is the associated Nambu-Goldstone boson which
could pick up mass in the presence of small terms in the scalar
potential with explicit \lnv that might arise, say, from quantum
gravity effects~\cite{kallosh:1995hi}.
Apart from its intrinsic interest as an attractive neutrino mass
generation scheme, it has been suggested that, the resulting
Nambu-Goldstone boson could play an interesting role in cosmology,
either providing a viable dark matter
candidate~\cite{berezinsky:1993fm,Lattanzi:2007ux,Lattanzi:2013uza,Queiroz:2014yna},
or bringing in the CP-even physical degree of freedom the state
responsible for driving
inflation~\cite{Boucenna:2014uma,Higaki:2014dwa}, as hinted by recent
cosmological data~\cite{Ade:2014xna}.

Besides providing a compelling framework for neutrino masses and
mixings, with all its phenomenological features, such majoron scenario
involves new ingredients which can affect the generation of the $B-L$
asymmetry. Let us discuss this in more detail. In type-I seesaw with
explicit lepton number violation, a nonzero $B-L$ asymmetry is
generated by the out-of-thermal-equilibrium and CP violating decays of
the lightest right-handed
neutrino~\cite{fukugita:1986hr,davidson:2008bu,Fong:2013wr}. This
$B-L$ asymmetry is then partially reprocessed into a baryon asymmetry
by standard model electroweak sphaleron processes
\cite{kuzmin:1985mm}, thus providing an explanation for the origin of
the cosmic baryon asymmetry. In this framework the RH neutrinos only
have Yukawa interactions, which are responsible for a non-vanishing RH
neutrino distribution as well as for the crucial $B-L$ washout
processes.  This picture is to some extent oversimplified and can
dramatically change whenever the RH neutrinos possess additional
interactions beyond those of the simplest type-I seesaw scenario.
This is indeed what happens in seesaw models with spontaneous lepton
number breaking in which RH neutrino masses rather than being put in
``by hand'', arise via a dynamical mechanism, similar to the breaking
of the standard model gauge symmetry.

In this paper we study the effects induced by these new right-handed
neutrino couplings upon the RH neutrino distribution, which ultimately
affect the resulting $B-L$ asymmetry. While similar considerations
apply also to theories of gauged lepton number, such as models with a
local U(1) lepton number symmetry \cite{valle:1987sq},
Pati-Salam~\cite{Pati:1974yy}, left-right
symmetric~\cite{mohapatra:1980ia}, here we focus upon the simplest \SM
scheme based upon spontaneously broken ungauged lepton
number~\cite{chikashige:1981ui,Schechter:1981cv}.
In addition to the standard model fields and three RH neutrinos, the
simplest of such seesaw models involves also a complex \SM singlet
scalar carrying two units of lepton number, denoted by $\sigma$.
The relevant invariant Yukawa interactions are
\begin{equation}
  \label{yukI}
  - \mathcal{L}_Y = \bar N_\alpha\,\lambda_{\alpha i}\,\ell_i\,\tilde H^\dagger
  + \frac{1}{2} \,\bar N_\alpha\,C\,h_{\alpha\beta} \,\bar N_\beta^T\,\sigma
  + \mbox{H.c.}\ ,
\end{equation}
where $\ell$ denotes the lepton doublet, $\lambda$ is a general
complex $3 \times 3$ matrix and $h$ is a complex symmetric $3 \times
3$ matrix in generation space. The resulting seesaw scheme is
characterized by singlet and doublet neutrino mass terms, described in
matrix form as
\begin{equation}
\label{ss-matrix} 
{\mathcal M_\nu} = 
\begin{pmatrix}
    0 & \lambda \vev{H} \\[+1mm]
    \lambda^{T} \vev{H}  & h \vev{\sigma} \\
\end{pmatrix} \, .
\end{equation}
Here $\vev{H}$ determines the masses of the weak gauge bosons, the
$W^\pm$ and the $Z^0$, hence $\vev{H}=v/\sqrt{2}\simeq 174\,$~GeV,
while the spontaneous \lnv occurs at the scale $\vev{\sigma} =
u/\sqrt{2}$. This vacuum expectation value (vev) drives spontaneous
\lnv and induces the RH Majorana neutrino mass matrix $M_N = h
\vev{\sigma}$ together with residual Yukawa interactions from $NN
\sigma$ in Eq.~(\ref{yukI}).
The diagonalization of the neutrino mass matrix \(\mathcal{M_\nu}\)
proceeds through a unitary mixing matrix as ${U}^T {\mathcal M_\nu} \,
U = \mathrm{diag}(m_i,M_i)$~\cite{Schechter:1980gr}, yielding 6 mass
eigenstates, the three light neutrinos with masses $m_i \sim
\lambda^2 v^2/M_N$, and three heavy neutrinos.
The gauge invariant tree-level Higgs scalar potential associated to
the singlet and doublet scalar multiplets $\sigma$ and $H$ is a simple
extension of that which characterizes the standard model and is given
by
\begin{equation}
  \label{eq:Vscalar}
  V(H,\sigma) = V_\text{SM}(H) + V_\text{BSM}(H,\sigma)=
  \lambda_H |H|^4 - m_H^2 |H|^2 
  + \lambda_\sigma |\sigma|^4 - m_\sigma^2 |\sigma|^2 
  + \delta |H|^2 |\sigma|^2 \ ,
\end{equation}
where the first two terms correspond to $V_\text{SM}$ and the
remaining ones account for either pure $\sigma$ interactions or the
singlet-doublet coupling. Note that cubic terms of the type $\sigma
|H|^2$ or $\sigma^3$ are absent due to lepton number conservation. The
full scalar potential $V$ will be bounded from below as long as the
conditions $\lambda_\sigma , \lambda_H > 0$ and $\delta > - 2
\sqrt{\lambda_\sigma \lambda_H}$ are satisfied.
After minimization, one finds, as expected, two physical CP-even
scalar bosons and one CP-odd state, identified with the majoron
$J$~\cite{chikashige:1981ui,Schechter:1981cv}, the Goldstone boson
associated to the spontaneous breaking of lepton number.

It is clear that in such schemes there will be a number of new
processes involving the new spin zero states which may have an impact
on the way the $B-L$ generation mechanism proceeds. In particular, the
RH neutrino Yukawa coupling $N\,N\,\sigma$ will induce $s,t,u$-channel
$2\leftrightarrow 2$ scattering processes $N_1 N_1\leftrightarrow
h_ih_j$ and $N_1 N_1\leftrightarrow JJ$. If the thermal bath is
populated with the new scalars before the leptogenesis era
\cite{Chung:1998rq}, these processes can efficiently populate the
plasma with RH neutrinos, basically resembling what happens in scalar
and fermion triplet leptogenesis models
\cite{Hambye:2005tk,Hambye:2012fh,AristizabalSierra:2010mv,Sierra:2014tqa}. If
this turns out to be the case, soon after these processes become
active the heat bath can be readily populated with a RH neutrino
thermal distribution, thus implying that if at early times the
$2\leftrightarrow 2$ scattering processes and the RH neutrinos inverse
decays are frozen, a sizeable enhancement of the $B-L$ yield can be
expected. Moreover, with these $2\leftrightarrow 2$ scattering
processes being fast at early epochs, the $B-L$ asymmetry will be
certainly independent upon initial conditions, even in the weak
washout regime. On the other hand, if scatterings remain efficient
till late epochs, the number of RH neutrinos available in the heat
bath can drastically decrease, thus leading to an important depletion
of the $B-L$ yield \cite{Gu:2009hn}. Our aim in this paper is to
provide a quantitative analysis of all these effects, identifying the
implications of these new reactions in the type-I seesaw leptogenesis
picture. In particular we explore up to which extent the condition of
successful leptogenesis constrains the parameters of these models.
Before we proceed let us emphasize the generality of our
study. Although motivated by type-I seesaw models with spontaneous
lepton number violation, these effects in seesaw leptogenesis are not
restricted to majoron models. In general, they will be present in all
models with extended scalar sectors that couple to the RH
neutrinos. These include all schemes with a $NN \sigma$ coupling, such
as other variants of majoron models~\cite{gelmini:1984ea}, flavon
models, etc.

The rest of the paper is organized as follows: in
Sec. \ref{sec:generalities} we establish the basic assumptions of our
analysis (fulfilled by the \lnv model we are interested in) and
present some general aspects of leptogenesis in models with extended
scalar sectors. In Sec. \ref{sec:BmL} we show our general results, to
be interpreted in the minimal majoron model in
Sec. \ref{sec:spontaneous-lepton-number-breaking}. Finally, we
summarize our results and conclude in Sec. \ref{sec:conclusions}.

\section{Generalities}
\label{sec:generalities}

All majoron schemes require an extended set of physical scalars.  We
now proceed to a more detailed and general study of the implications
of having additional scalars in leptogenesis seesaw scenarios. The
basic particle physics input is the CP asymmetry parameter
$\epsilon_{N_1}$ characteristic of seesaw
leptogenesis~\cite{fukugita:1986hr,davidson:2008bu,Fong:2013wr}.
In order to determine the final $B-L$ asymmetry yield one must also
take into account all $B-L$ production and washout processes.
In general, the scalar sector gives rise to new contributions to
these. For example, the new scalars could contribute to the CP
asymmetry $\epsilon_{N_1}$ as well as provide new source/washout terms
in the kinetic equations, potentially affecting the constraints on the
$B-L$ generation scale~\cite{Davidson:2002qv}.
However, we neglect such possible contributions and focus on the role
of the RH neutrino scatterings, assuming the following conditions:
\begin{enumerate}[$(A)$]
\item \label{item:H4} In addition to the usual type-I seesaw
  Lagrangian (in the basis where the RH neutrino mass matrix is
  diagonal)~\footnote{We choose to denote RH neutrino generations with
    Greek letters $\alpha, \beta, \gamma, \dots$, lepton flavors with
    the Latin letters $i, j, k, \dots$, while the extra scalars are
    labeled with the Latin letters $a, b, c, \dots$.}
  \begin{align}
  \label{eq:seesaw-interactions}
  {\cal L}_\text{seesaw} = 
  \bar N_\alpha\,\lambda_{\alpha i}\,\ell_i\,\tilde H^\dagger
  + \frac{1}{2}\bar N_\alpha\,C\,M_{N_{\alpha\alpha}}\,\bar N_\alpha^T
  + \mbox{H.c.}\ ,
\end{align}
there are new interactions generically described by the following
terms
  \begin{align}
  \label{eq:interactions}
  {\cal L} \supset
  \frac{1}{2}g_{N_\alpha N_\beta a}\,\bar N_\alpha\,C\,\bar N_\beta^T\,S_a
  + \mu_{a b c}\,S_a\,S_b\,S_c\  + \mbox{H.c.}
\end{align}
The dimensionless coupling $g_{N_\alpha N_\beta a}$ and the mass
parameters $\mu_{a b c}$ characterize the scalar sector as arising,
for example, in the type-I seesaw majoron scheme. Here the $S_a$'s
correspond to mass eigenstates.
\item \label{item:H1} All the CP violating phases involved in the
  generation of a non-vanishing $B-L$ asymmetry are entirely
  attached to the Dirac Yukawa couplings.
\end{enumerate}
Under these assumptions the leading ${\cal O}(\lambda^2)$ reactions
relevant for the generation of the $B-L$ asymmetry include the
following processes: $\Delta L=1$ decay and inverse decay reactions
$N_1\leftrightarrow \ell\,H$; $s$-channel off-shell $\Delta L=2$
scatterings $H\,\ell\leftrightarrow H\,\ell$ and $s,t,u$-channel $N_1
N_1\leftrightarrow S_a\,S_b$ scatterings implied by
Eq.~(\ref{eq:interactions}).
The new $2\leftrightarrow 2$ scattering reactions resemble those found
in type-II or type-III seesaw leptogenesis scenarios, where the states
responsible for the generation of the $B-L$ asymmetry--having
non-trivial electroweak charges--possess vector-boson-mediated
annihilations \cite{Hambye:2003rt,Hambye:2005tk,Hambye:2012fh,
  AristizabalSierra:2010mv,AristizabalSierra:2011ab,Sierra:2014tqa}.
These reactions are also present when RH neutrinos have non-trivial
gauge charges, as in left-right symmetric models
\cite{Cosme:2004xs,Frere:2008ct,Blanchet:2010kw} or models with
additional $U(1)$ gauge groups
\cite{Plumacher:1996kc,Racker:2008hp,Blanchet:2009bu,Buchmuller:2012wn}.
However, compared with majoron models these scenarios differ in
several aspects. For example: $(i)$ in some cases the resulting $B-L$
asymmetry can be a combination of non-thermal and thermal
contributions \cite{Buchmuller:2012wn}; $(ii)$ the gauge scattering
processes, although present, are of no relevance
\cite{Blanchet:2010kw}. Furthermore, there is a fundamental
difference: while the vector-boson-mediated annihilations at high
temperatures are always fast (their rate is faster then the Universe
Hubble expansion rate), the RH neutrino annihilations in majoron
models are controlled by free parameters. Depending on the size of the
couplings in Eq.~(\ref{eq:interactions}) and on the RH neutrino mass,
these scattering reactions may have new effects.
First, if after reheating, the heat bath turns out to be populated
with the new scalars (in addition to the standard model
particles)~\cite{Chung:1998rq}, the $2\leftrightarrow 2$ scatterings
could populate the plasma with RH neutrinos in addition to those
coming from RH neutrino inverse decays.
On the other hand, for sufficiently fast scattering processes the
resulting $B-L$ asymmetry will no longer depend upon initial
conditions, even if the asymmetry is generated within the weak washout
regime. Moreover, fast $2\leftrightarrow 2$ processes will tend to
thermalize the RH neutrino distribution. This implies that the
generation of the $B-L$ asymmetry will no longer be determined solely
by reactions involving the Dirac Yukawa couplings $\lambda$ but
instead by an interplay between decays and RH neutrino annihilations.

Rather than demanding the scalar-mediated RH neutrino annihilations to
be slow, $\gamma_S/n^\text{Eq}_N\,H\lesssim 1$ ($\gamma_S$ being the
annihilation reaction density), what really matters is the relative
size between the decay and annihilation reaction densities. Thus, one
can distinguish two types of scenarios, those for which $\gamma_S\ll
\gamma_D$, $\gamma_D$ denoting the decay reaction density, and those
for which the annihilation processes are fast and satisfy $\gamma_S >
\gamma_D$ during certain range in $z=M_{N_1}/T$. In the former case
the annihilation reactions are negligible and the generation of the
$B-L$ asymmetry proceeds in the standard way. On the other hand in the
latter case a sizeable $B-L$ asymmetry requires decays and scatterings
to become slow.

\begin{figure}
  \centering
  \includegraphics[scale=0.33]{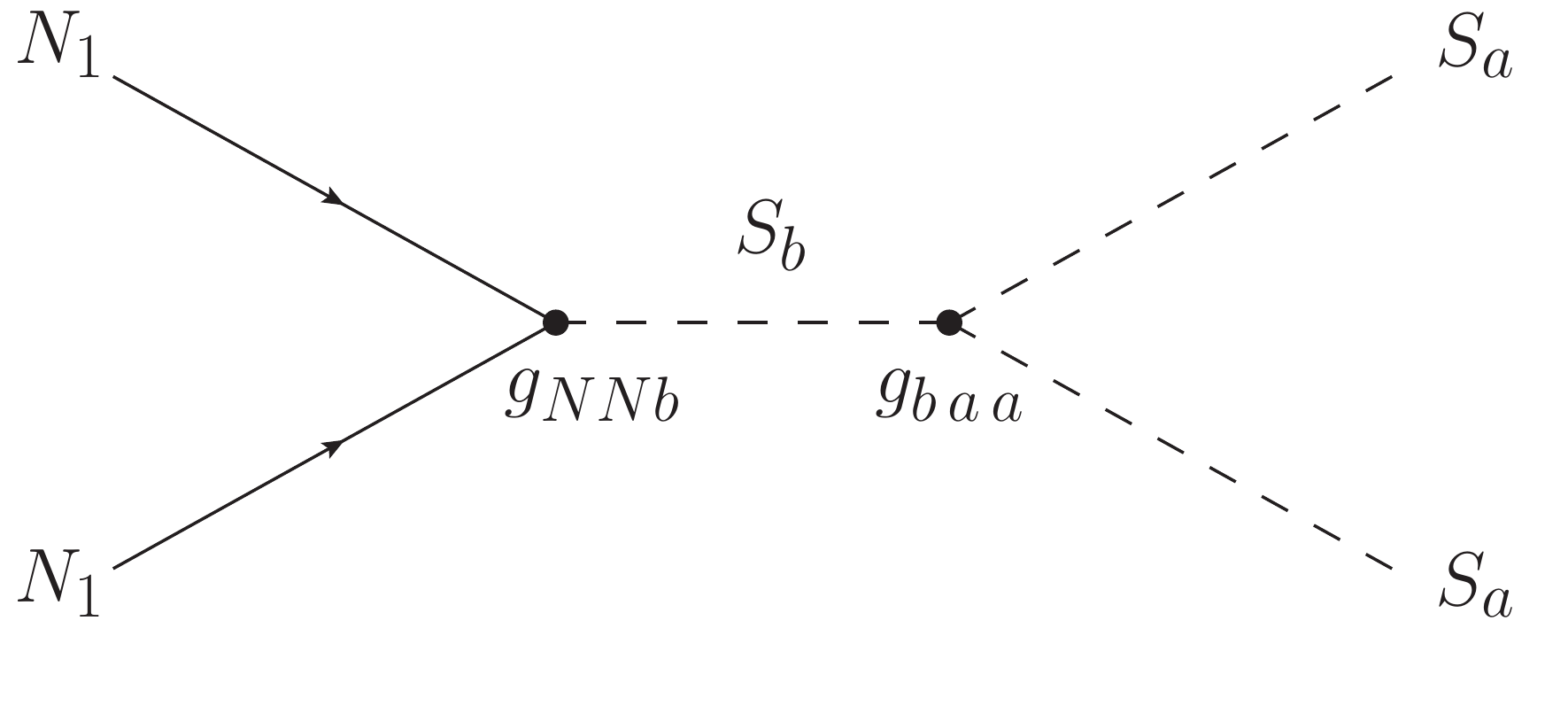}
  \hfill
  \includegraphics[scale=0.4]{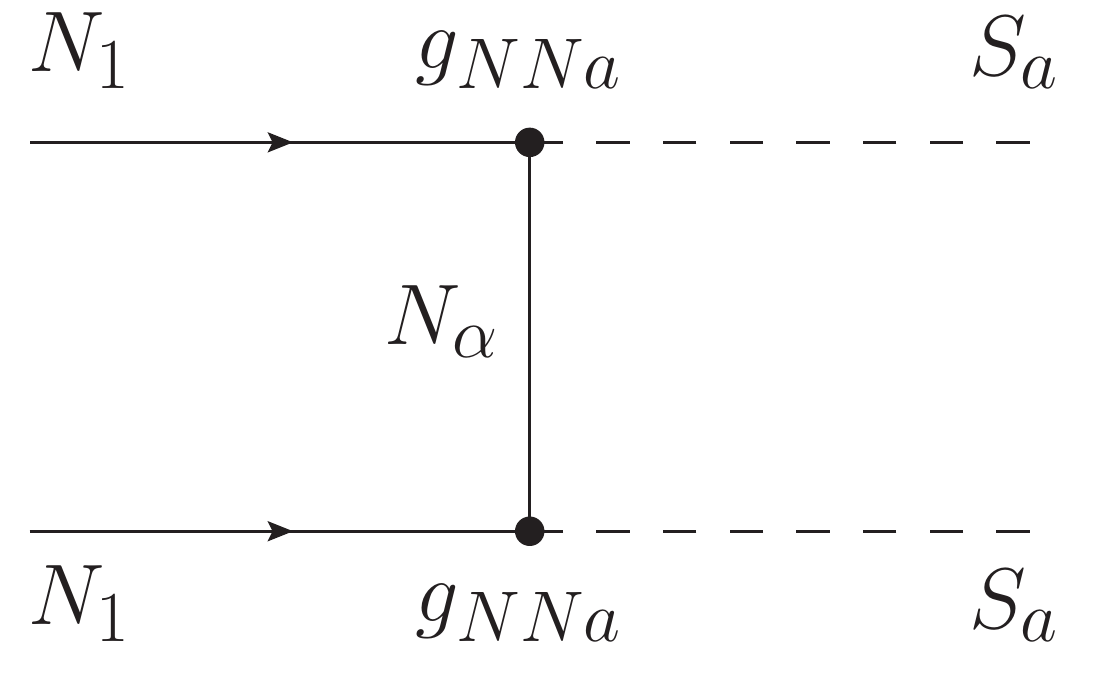}
  \hfill
  \includegraphics[scale=0.4]{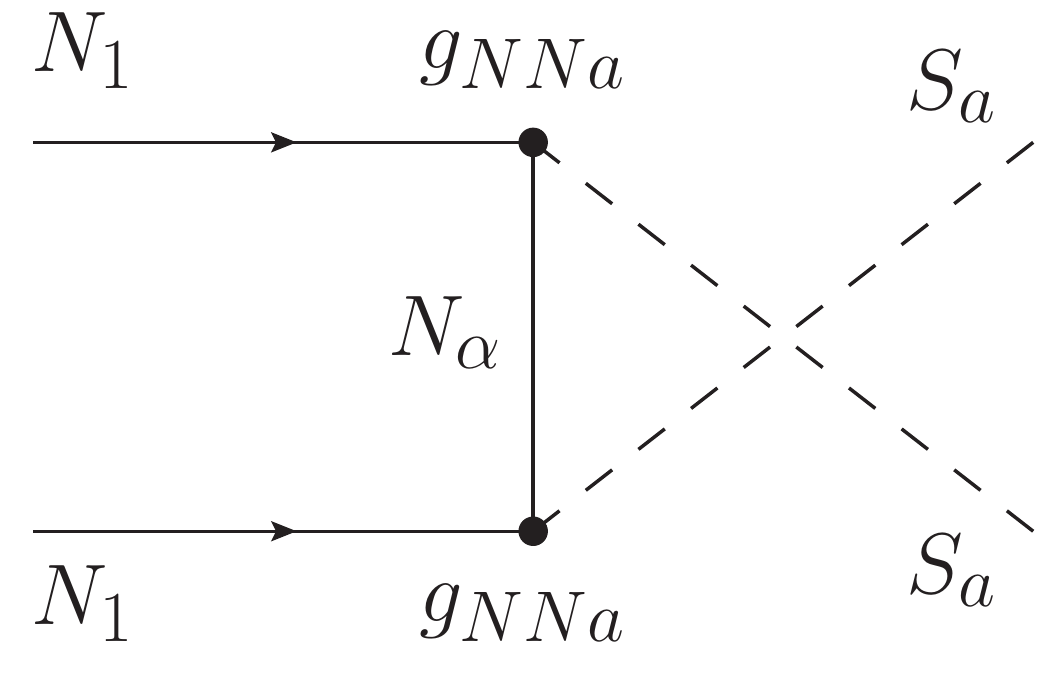}
  \caption{\it $s$-, $t$- and $u$-channel processes contributing to
    the total $N_1\,N_1\to S_a\,S_a$ annihilation cross section. In
    our calculations we will only consider diagrams with $N_\alpha =
    N_1$ (see footnote \ref{foot3}). }
  \label{fig:feynman-diag}
\end{figure}

In order to illustrate the general features of these scenarios one
must calculate the annihilation reaction density, and this requires
specifying the number of scalar degrees of freedom participating in
the annihilation process. For definiteness, here we assume that the
extended scalar sector contains three new scalars $S_1$, $S_2$ and
$S_3$ of which $S_{1,2}$ are CP even states and $S_3$ is a CP odd
field. This pattern arises in the simplest majoron scheme, in which
case $S_3$ would correspond to the majoron.
When combined with assumption \ref{item:H1}, these CP transformation
properties imply the absence of interactions with an odd power in
$S_3$, such as $N_\alpha\,N_\alpha\,S_3$ and $S_a\,S_b\,S_3$. Thus,
the presence of these new scalar states induces the scattering
processes $N_1\,N_1\to S_a\,S_a$ \footnote{Strictly speaking, in
  addition to CP conservation, we are also assuming that all RH
  neutrinos have the same CP
  parity~\cite{Schechter:1981hw,wolfenstein:1981rk}.  Moreover, for
  simplicity, we neglect off-diagonal scattering processes
  $N_1\,N_1\to S_a\,S_b$, with $a\neq b$, as they would not add any
  essentially new feature.}  determined by $s$-channel $S_{1,2}$
exchange and $t$- and $u$-channel $N_1$-mediated processes
\footnote{In principle one should also consider $N_{2,3}$-mediated
  processes. In order to reduce the number of free parameters we have
  neglected these contributions. This is well justified for strongly
  hierarchical RH neutrinos, where such contributions are
  sub-leading. \label{foot3}}, as shown in
Fig.~\ref{fig:feynman-diag}. As a function of the annihilation
center-of-mass energy $\sqrt{s}$, the cross section for the
$N_1\,N_1\to S_a S_a$ scattering processes is then given by four
terms, namely
\begin{equation}
  \label{eq:x-section-full}
  \sigma(N_1\,N_1\to S_a\,S_a)\equiv \sigma_S^{(a)}(s) = 
  \sigma_{s}^{a}(s)
  + \sigma_{s_1(t,u)}^{(a)}(s)
  + \sigma_{s_2(t,u)}^{(a)}(s)
  + \sigma_{(t,u)}^{(a)}(s)\ ,
\end{equation}
with the first term arising from the $S_{1,2}$ $s$-channel processes,
the second and third terms from the interference between $s$-channel
and $t$- and $u$-channel reactions, while the last term is due to
$t$- and $u$-channel processes only. Explicitly, we have
\begin{align}
  \label{eq:x-section-complete1}
  \sigma_s^{(a)}(s)&=\frac{1}{16\pi\,s}\frac{r_{S_a}}{r_N}\frac{\omega}{x_N}
  \left(
    \frac{g_{1aa}^2\;g_{NN1}^2}{\bar r_{S_1}^2}
    + 4\frac{g_{1aa}\;g_{2aa}\;g_{NN1}\;g_{NN2}}{\bar r_{S_1}\bar r_{S_2}}
    + \frac{g_{2aa}^2\;g_{NN2}^2}{\bar r_{S_2}^2}
  \right)\ ,
  \\
  \label{eq:x-section-complete2}
  \sigma_{s_{(1,2)}(t,u)}^{(a)}(s)&=\frac{g_{NN(1,2)}\;g_{NNa}^2
    \;g_{(1,2)aa}}{4\pi\,s}
  \frac{1}{r_N^2\,\bar r_{S_{(1,2)}}}\frac{\omega}{x_N}\;
  \log\left[\frac{c_{Na}\;x_N (1 - r_N\;r_{S_a}) - 2} {c_{Na}\;x_N (1
      + r_N\;r_{S_a}) - 2}\right]\ ,
  \\
  \label{eq:x-section-complete3}
  \sigma_{(t,u)}^{(a)}(s)&=\frac{g_{NNa}^4}{2\,\pi\,s}
  \frac{r_{S_a}}{r_N}\,c_{Na}\,\omega\,
  \left\{
    \frac{c_{Na}\,x_N}{\left[c_{Na}\,x_N(1 + r_N\,r_{S_a}) - 2\right]
      \left[c_{Na}\,x_N(1 - r_N\,r_{S_a}) - 2\right]}
  \right.
  \nonumber\\
  &\;\;\;\;\left. + \frac{1}{2r_N\;r_{S_a}\;(c_{Na}\;x_N - 2)}
    \log
    \left[
      \frac{c_{Na}\;x_N (1 + r_N\;r_{S_a}) - 2}
      {c_{Na}\;x_N (1 - r_N\;r_{S_a}) - 2}
    \right]
  \right\}\ .
\end{align}
In these expressions the following conventions have been adopted:
$r_N=\sqrt{1-4/x_N}$, $r_{S_a}=\sqrt{1-(4/c_{Na}\,x_N)}$, $\bar
r_{S_{1,2}}=1 - (1/c_{N(1,2)}\,x_N)$ and $\omega=1-2/x_N$, with
$x_N=s/M_{N_1}^2$ and $c_{Na}=M_{N_1}^2/m_{S_a}^2$. Moreover, we
denote $g_{N N a} \equiv g_{N_1 N_1 a}$. Under our assumptions the
left diagram can produce three scalars whereas the others only produce
$S_1$ and $S_2$.  We have also introduced the dimensionless parameter
$g_{(1,2)aa}=\mu_{(1,2)aa}/M_{N_1}$. In the derivation of the cross
section none of the scalar masses have been neglected, the vanishing
scalar mass limit--applicable in a large variety of
scenarios--corresponds to $c_{Na}\to \infty$. The total cross section
is then given by
\begin{equation}
  \label{eq:total-x-section-including-sum_a}
  \sigma_S(s)=\sum_{a=1}^3\sigma^{(a)}_S(s)\ ,
\end{equation}
Using the expressions Eq.~(\ref{eq:x-section-complete1}),
(\ref{eq:x-section-complete2}) and (\ref{eq:x-section-complete3}) one
can derive the reduced cross section,
$\widehat\sigma(x_N)=2s\,\sigma_S(s)\,r_N^2$, that enables the
calculation of the $N_1\,N_1\to S_a\,S_a$ reaction density:
\begin{equation}
  \label{eq:annihilation-reaction-density}
  \gamma_S = \frac{M_{N_1}^4}{64\,\pi^4}\,\int_4^\infty\,dx_N\,\sqrt{x_N}\,
  \frac{K_1(z\sqrt{x_N})}{z}\,\widehat\sigma_S(x_N)\ ,
\end{equation}
where $K_1(z)$ is a modified Bessel function.  For the decay reaction
density, instead, the following expression holds
\begin{equation}
  \label{eq:decay-RD}
  \gamma_D=\frac{M_{N_1}^3}{\pi^2}\frac{K_1(z)}{z}\,\Gamma_{N_1}^\text{Tot}
  \qquad\text{with}\qquad
  \Gamma_{N_1}^\text{Tot}=\frac{M_{N_1}^2}{8\;\pi\;v^2}\;\widetilde m_1\ ,
\end{equation}
with $v$ defined as in Sec. \ref{sec:intro}, namely
$\vev{H}=v/\sqrt{2}\simeq 174\,$~GeV, and the ``effective'' mass
parameter $\widetilde m_1$, determined by the leading seesaw
contribution to light neutrino masses, 
\begin{equation}
  \label{eq:mtilde}
  \widetilde m_1=\frac{v^2}{M_{N_1}}
  \,\left(\lambda\,\lambda^\dagger\right)_{11}\ .
\end{equation}
With the setup of
Eqs. (\ref{eq:x-section-complete1})-(\ref{eq:x-section-complete3}),
(\ref{eq:annihilation-reaction-density}) and (\ref{eq:decay-RD}), the
different scenarios one can consider can now be illustrated. In order
to proceed one must specify a point in parameter space which consist
of $\tilde m_1$, $M_{N_1}$, the scalar and RH neutrino mass
hierarchies, $c_{Na}$, and the dimensionless couplings $g_{(1,2)aa}$ and
$g_{NNa}$. However, even without sticking to particular parameter
values, general conclusions can be made by noting that the relative
size of the annihilation and decay reaction densities is mainly
determined by 
\begin{equation}
  \label{eq:gammaS-over-gammaD}
  \frac{\gamma_S}{\gamma_D}\sim \frac{\bar g^4v^2}
  {M_{N_1}\widetilde m_1}\ ,
\end{equation}
where with $\bar g$ we refer to any of the couplings entering in the
cross section in
Eqs.~(\ref{eq:x-section-complete1})-(\ref{eq:x-section-complete3})~\footnote{Note
  that $\bar g$ denotes any of the parameters involved in the
  scattering cross section, which only involves $N_1$ (and not
  $N_{2,3}$). This is because, by assumption (see footnote
  \ref{foot3}), the dynamics of the problem is fully dictated by the
  lightest RH neutrino, assumed to be much lighter than the other
  two. For this reason, all parameters in
  Eq.\eqref{eq:gammaS-over-gammaD} are related only to $N_1$.}. For
simplicity, in most of our analysis we will assume ``universality'' of
these couplings.
Thus, for a given $M_{N_1}$, the relevance of the scalar-induced RH
neutrino scattering will depend on $\widetilde m_1$ and $\bar g$. A
small $\bar g$ will lead to a sub-dominant scattering process
($\gamma_S<\gamma_D$) unless $\widetilde m_1$ is small as well (the
precise values determined by specific parameter choices), while large
values of $\bar g$ will render the RH neutrino scattering a dominant
process up to large values of $z$. As shown in the following section,
this behavior has striking consequences, namely it can either
dramatically enhance or suppress the $B-L$ yield.

\begin{figure}
  \centering
  \includegraphics[scale=0.65]{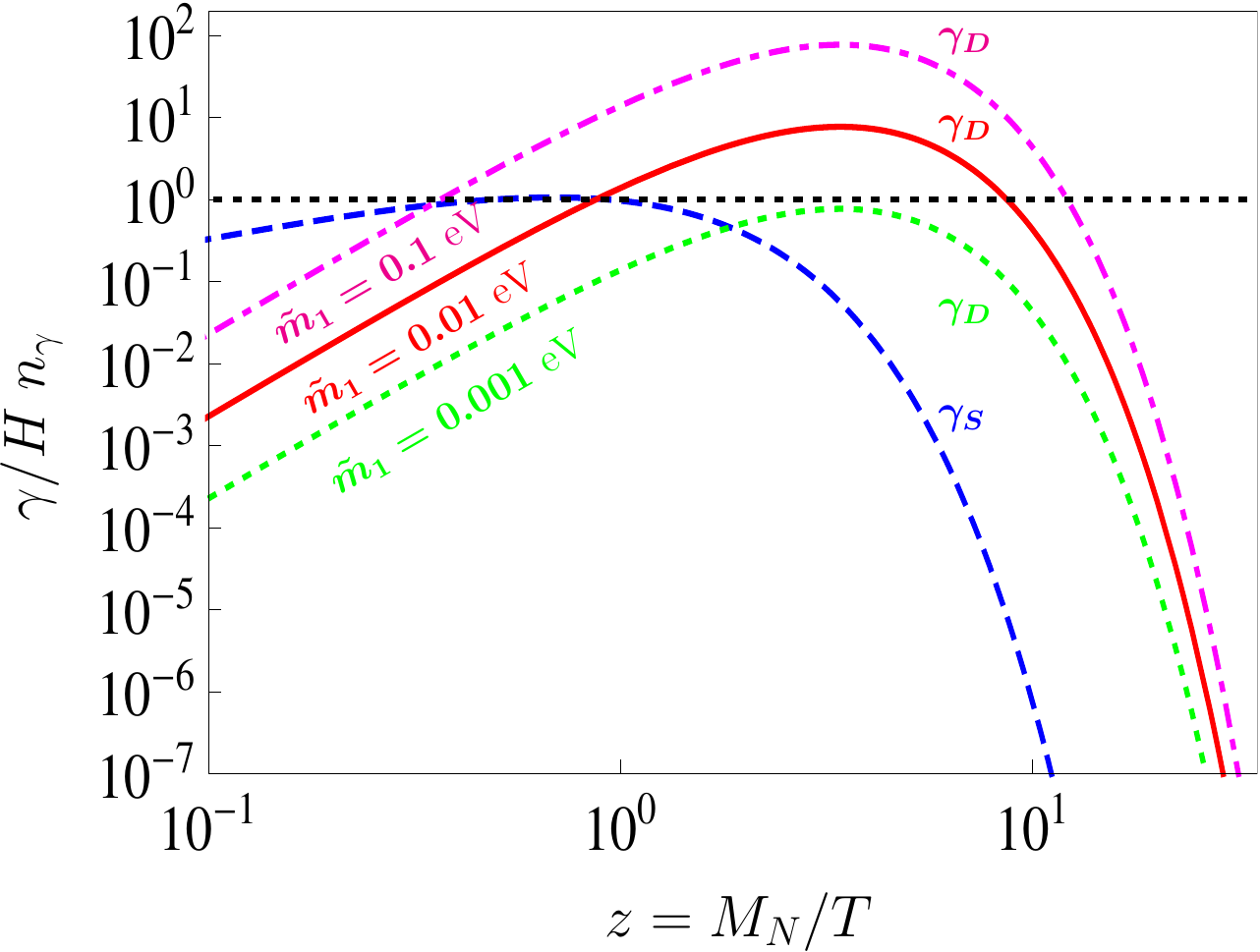}
  \hfill
  \includegraphics[scale=0.65]{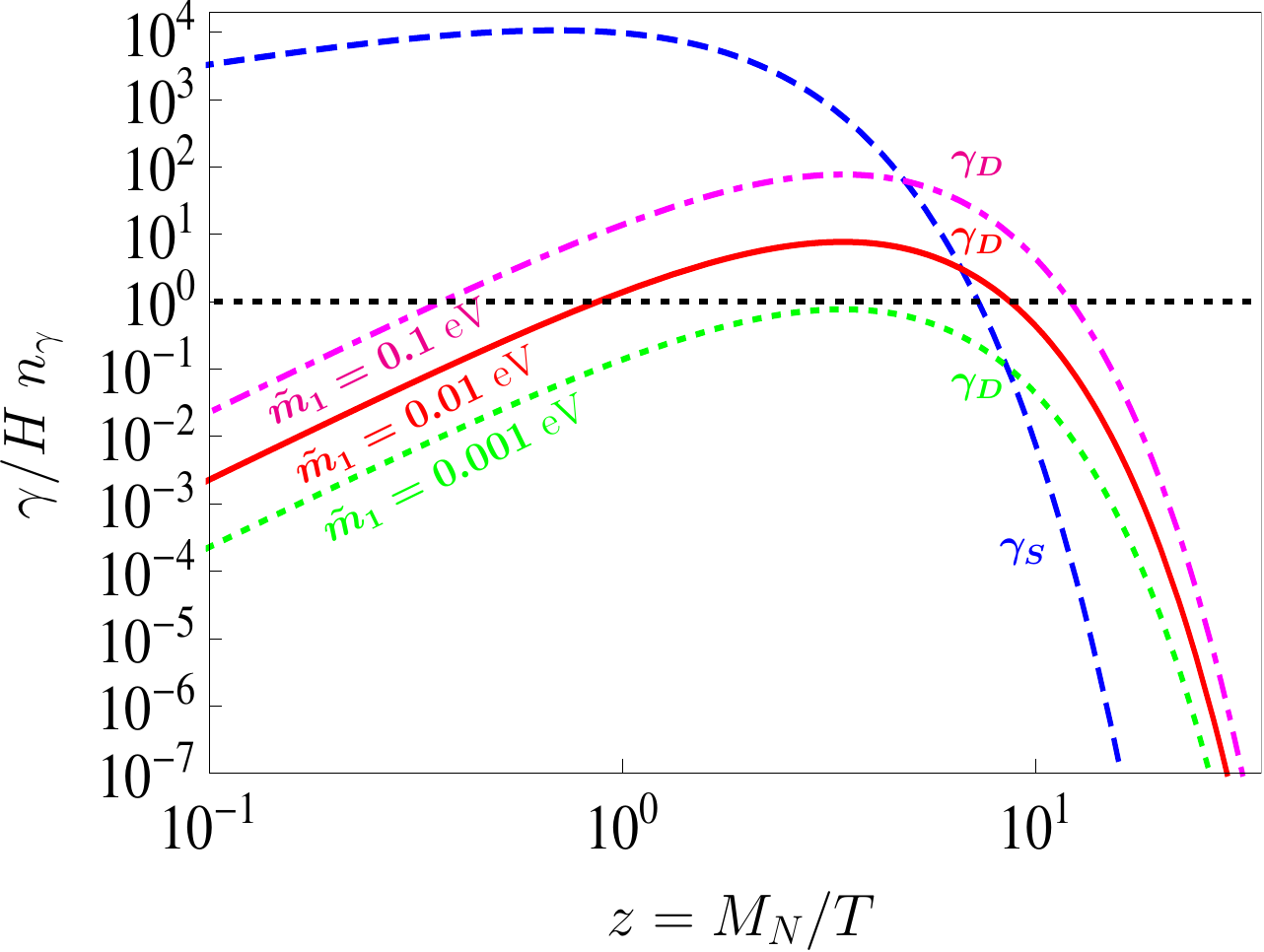}
  \caption{Right-handed neutrino scalar-mediated annihilation and
    decay reaction densities as a function of $z=M_{N_1}/T$ for
    $M_{N_1}=10^{10}$~GeV. In the left plot we have chosen $\bar g =
    10^{-1}$, whereas in the right plot $\bar g = 1$. See text for
    details.}
  \label{fig:reaction-densities}
\end{figure}
Although there is no fundamental reason for the different
dimensionless couplings to be equal nor for the scalar masses to be
well below the RH neutrino mass, it turns out that these simplifying
``universality'' assumptions capture the main features of the role
played by RH neutrino scatterings in seesaw leptogenesis (see
discussion in section \ref{sec:beyond-simpl-scen}). The generic
behavior described above is illustrated in
Fig.~\ref{fig:reaction-densities}, where for concreteness we have
taken the RH neutrino mass to be $10^{10}$~GeV, and the dimensionless
couplings to be $10^{-1}$ (left panel) and $1$ (right panel). So, for
example, while a decay controlled by $\widetilde m_1=0.01$~eV becomes
dominant already at $z\sim 1$ when $\bar g=10^{-1}$, (i.e. RH
neutrinos will not be able to thermalize by the scatterings), the same
decay will be severely swamped by RH neutrino scatterings when $\bar
g=1$.

\section{The baryon asymmetry from leptogenesis}
\label{sec:BmL}

With the general picture already clear, in this section we will
analyze the implications of the scalar-mediated right-handed neutrino
scatterings for the generation of the $B-L$ asymmetry. For this aim we
need to write suitable kinetic equations with which the $B-L$
asymmetry can be tracked. Depending on the dynamics of the new
scalars, these equations can substantially differ from those of the
standard case~\cite{Giudice:2003jh,Nardi:2007jp}. 

\subsection{Dynamical features of the $B-L$ asymmetry generation}

First note that if the heat bath before the leptogenesis era becomes
populated with the new scalars, and their interactions are such that
their distributions obey
\begin{equation}
  \label{eq:scalar-densities}
  \frac{Y_{S_a}}{Y_{S_a}^\text{Eq}}\simeq 1 + \theta 
  \qquad\text{with}\qquad \theta\ll 1\ ,
\end{equation}
then the corresponding kinetic equations take a rather simple form,
similar to the kinetic equations found in fermionic triplet
leptogenesis scenarios
\cite{Hambye:2003rt,AristizabalSierra:2010mv,Hambye:2012fh}. Note that
the new scalars will at least couple via scalar vertices to the
standard model Higgs doublet~\footnote{Depending on the model the new
  scalars might also have gauge and/or new Yukawa interactions.}. The
thermal bath will then be populated with the new scalars either
directly after reheating \cite{Chung:1998rq}, or through the
interactions characterizing the scalar sector in
Eq.~(\ref{eq:interactions}). In this case
Eq.~(\ref{eq:scalar-densities}) will hold provided the scalar
interactions are strong enough.
Using Eq.~(\ref{eq:scalar-densities}) we write the system of coupled
Boltzmann differential equations at order $\lambda^2$ from
Eq.~(\ref{eq:seesaw-interactions})
as~\footnote{We neglect spectator processes, work in the
  one-flavor-approximation, and drop order $\theta$ terms.}
\begin{align}
  \label{eq:BEQs-YN}
  \frac{d}{dz}Y_N&=-\frac{1}{\mathsf{s}Hz}
  \left\{
    \left(
      \frac{Y_N}{Y_N^\text{Eq}} - 1
    \right)\gamma_D
    +
    \left[
      \left(
        \frac{Y_N}{Y_N^\text{Eq}}
      \right)^2 - 1
    \right]\gamma_S
  \right\}\ ,
  \\
  \label{eq:BEQs-YDBmL}
  \frac{d}{dz}Y_{\Delta_{B-L}}&=-\frac{1}{\mathsf{s}Hz}
  \left[
    \left(
      \frac{Y_N}{Y_N^\text{Eq}} - 1
    \right)\epsilon_N
    +
    \frac{Y_{\Delta_{B-L}}}{2\,Y^\text{Eq}_\ell}
  \right]\gamma_D\ .
\end{align}
\begin{figure}
\centering
\includegraphics[scale=0.65]{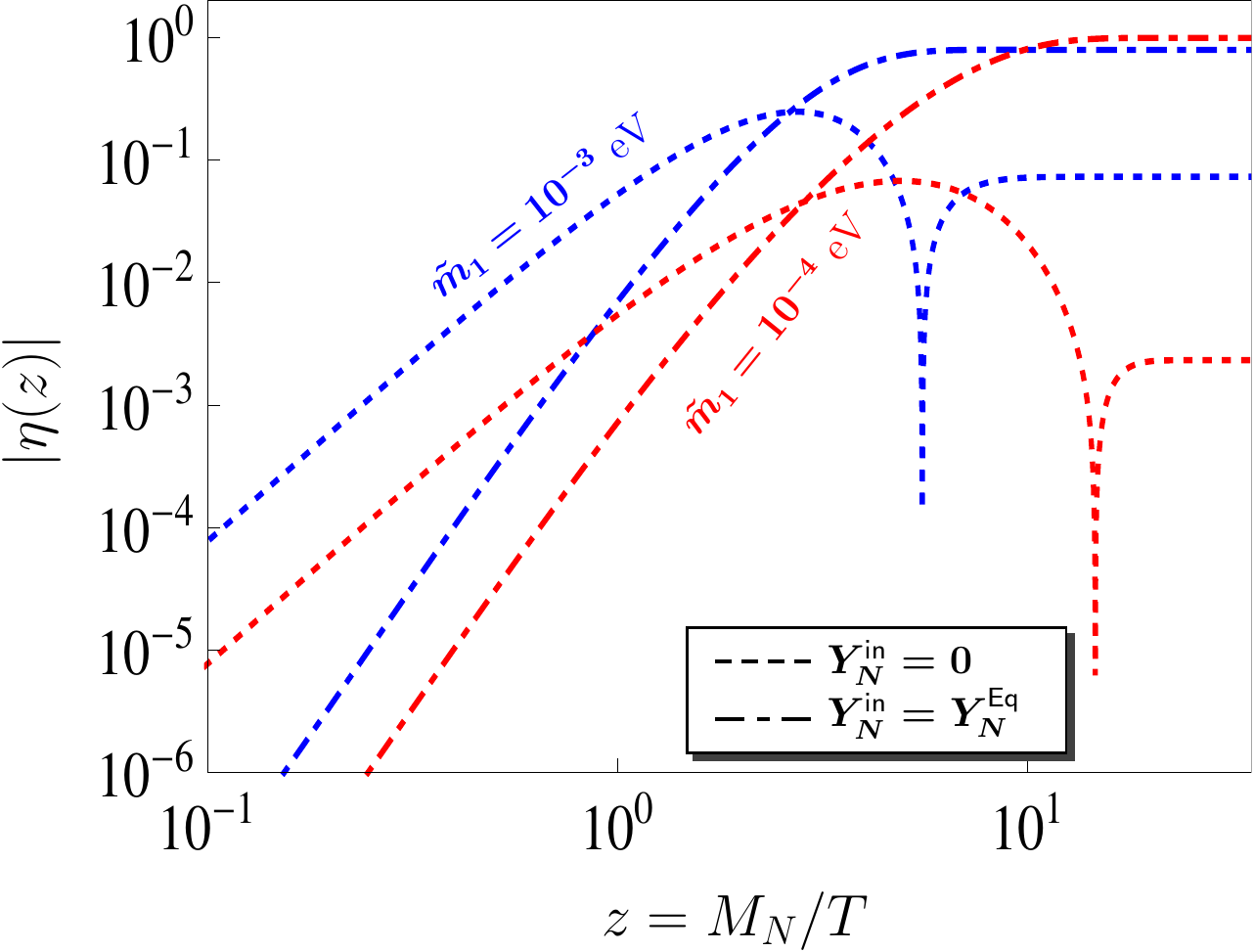}
\hfill
\includegraphics[scale=0.65]{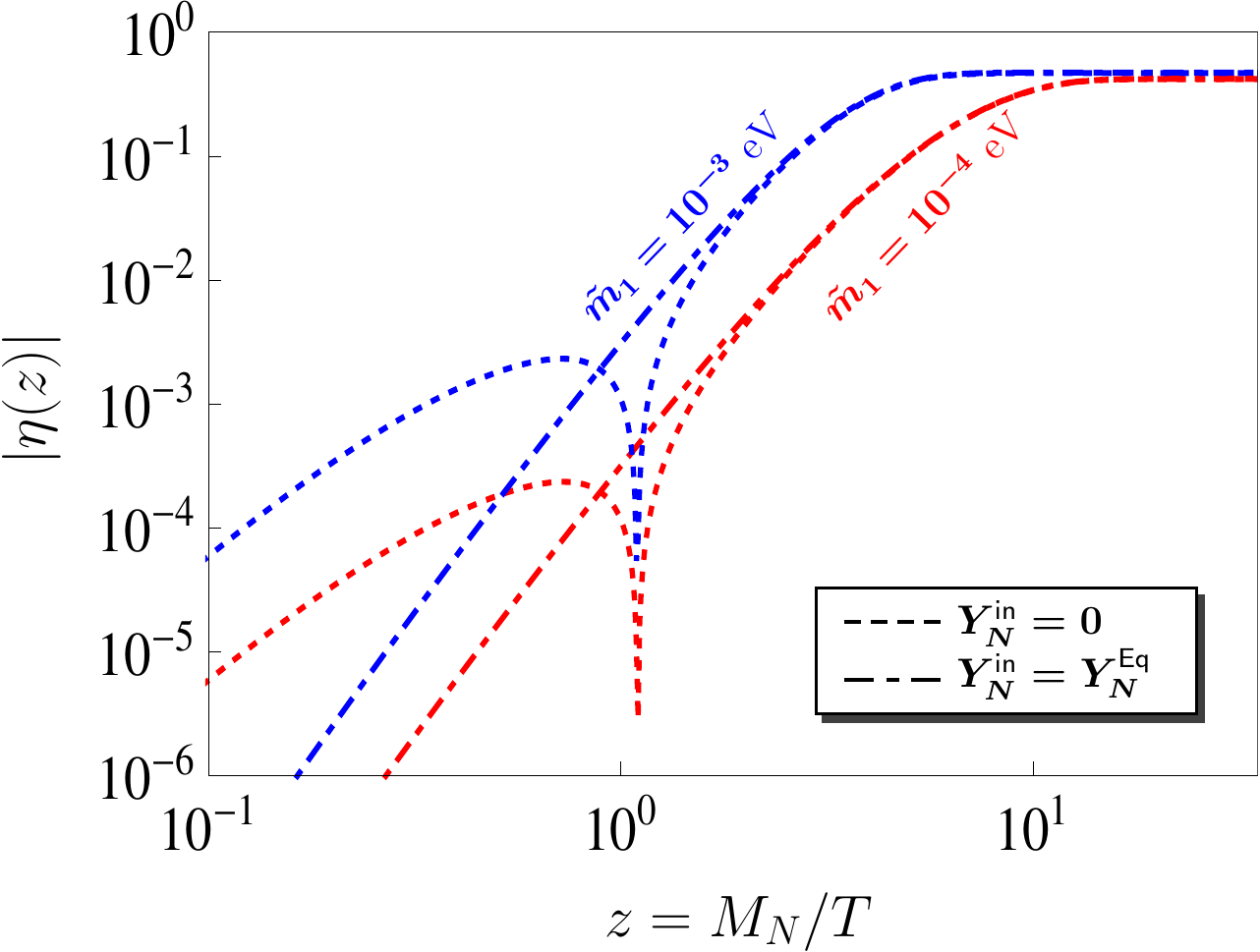}  
\caption{Efficiency function $\eta(z)$ in Eq.~(\ref{eq:BmL-formal-exp})
  for two representative $\widetilde m_1$ values in the weak washout
  regime. The left panel is the standard case, while the right panel
  includes the effects of right-handed scatterings with $\bar g =
  10^{-1}$ and $M_{N_1}=10^{10}$~GeV. One sees that the $B-L$
  asymmetry at the end of the leptogenesis era is independent of the
  initial RH neutrino distribution.}
  \label{fig:eff-vs-z}
\end{figure}
Here $Y_X = n_X/\mathsf{s}$ is the number density-to-entropy ratio of
species $X$ and $\mathsf{s}$ denotes the entropy density.  Formal
integration of these equations allows one to express the $B-L$
asymmetry in terms of the efficiency $\eta$ as~\cite{Barbieri:1999ma}
\begin{equation}
  \label{eq:BmL-formal-exp}
  Y_{\Delta_{B-L}}(z)=Y^\text{Eq}_N(z\to 0)\,\epsilon_N\,\eta(z)\ ,
\end{equation}
with the final $B-L$ yield $Y_{\Delta_{B-L}}(z\to \infty)$ determined
by the particle physics CP asymmetry parameter $\epsilon_N$ and the
washout dynamics, characterized by the efficiency parameter $\eta$
defined as $\eta\equiv\eta(z\to\infty)$. Here $Y^\text{Eq}_N(z\to 0)$
is a normalization factor.

Integrating equations (\ref{eq:BEQs-YN}) and (\ref{eq:BEQs-YDBmL}) we
now derive the consequences of the new RH neutrino scattering
reactions on the generation of the $B-L$ asymmetry.
We first consider the weak washout regime, given by the condition
$\widetilde m_1\lesssim m_\star$ with $m_\star\simeq 10^{-3}$~eV,
entirely determined by cosmological input, see
e.g. \cite{davidson:2008bu}. We calculate the efficiency for two
values of $\widetilde m_1=10^{-3}, 10^{-4}$~eV assuming two different
initial RH neutrino densities: vanishing and equilibrium, and for the
parameters fixed as in the left panel in
Fig.~\ref{fig:reaction-densities}.
As seen in the left panel in Fig.~\ref{fig:eff-vs-z}, in the standard
case the $B-L$ yield strongly depends upon the initial RH neutrino
abundance, as expected~\cite{Barbieri:1999ma}.
In contrast, as we have pointed out in the previous section, the
presence of the new scattering processes can change that picture
drastically provided the scattering reaction rate for $S_a\,S_a\to
N_1\,N_1$ is larger than the inverse decay rate at early times (high
temperatures). In such case the heat bath will be populated with RH
neutrinos through scattering processes rather than by slow inverse
decays.  This result clearly demonstrates that the presence of the
scalar-induced RH neutrino scattering renders the $B-L$ yield
independent of assumptions regarding the initial RH neutrino
population, involving only $\epsilon_N$, $\tilde m_1$, $M_{N_1}$ and
$\bar g$, see right panel in Fig.~\ref{fig:eff-vs-z}.

\begin{figure}
  \centering
  \includegraphics[scale=0.65]{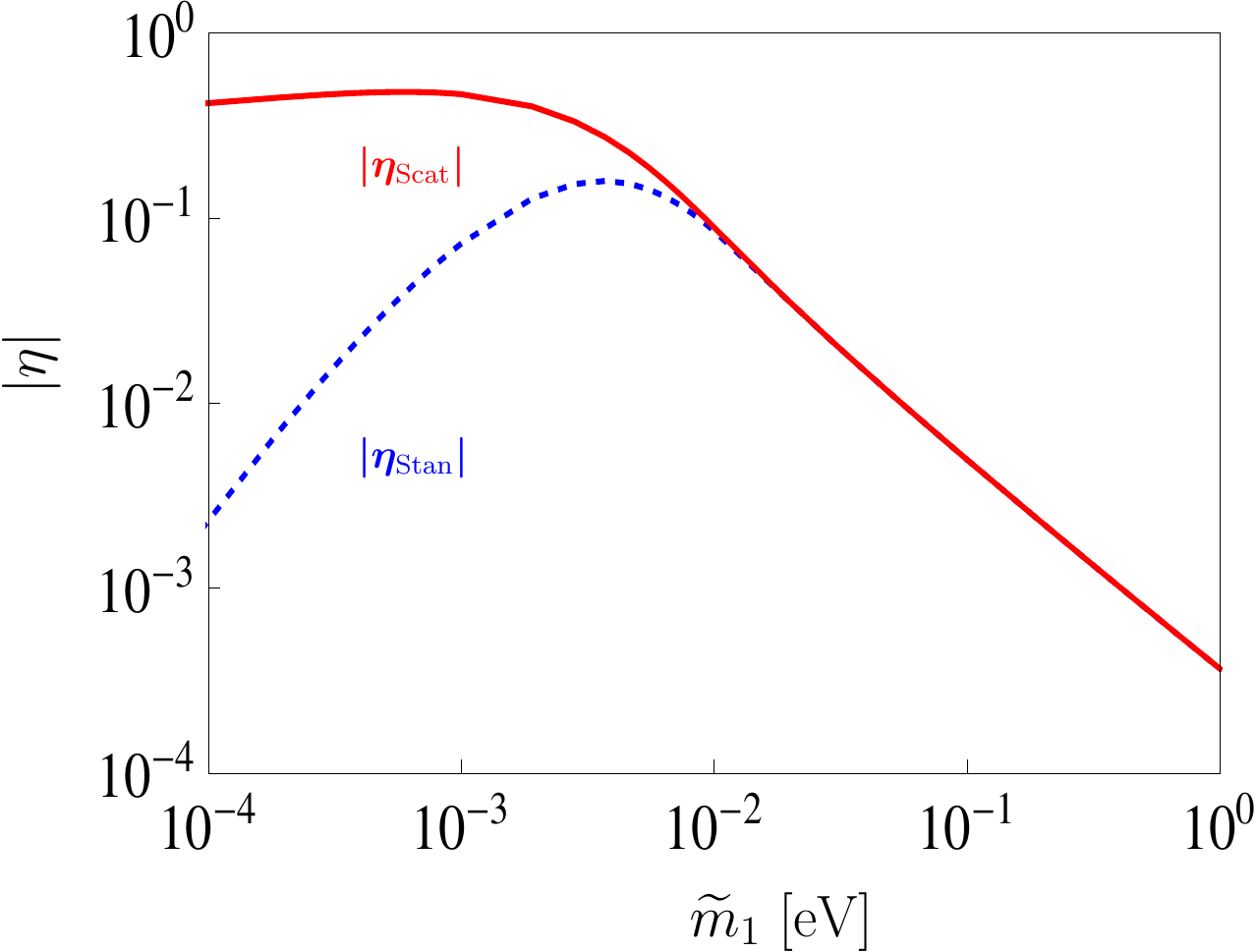}
  \hfill
  \includegraphics[scale=0.65]{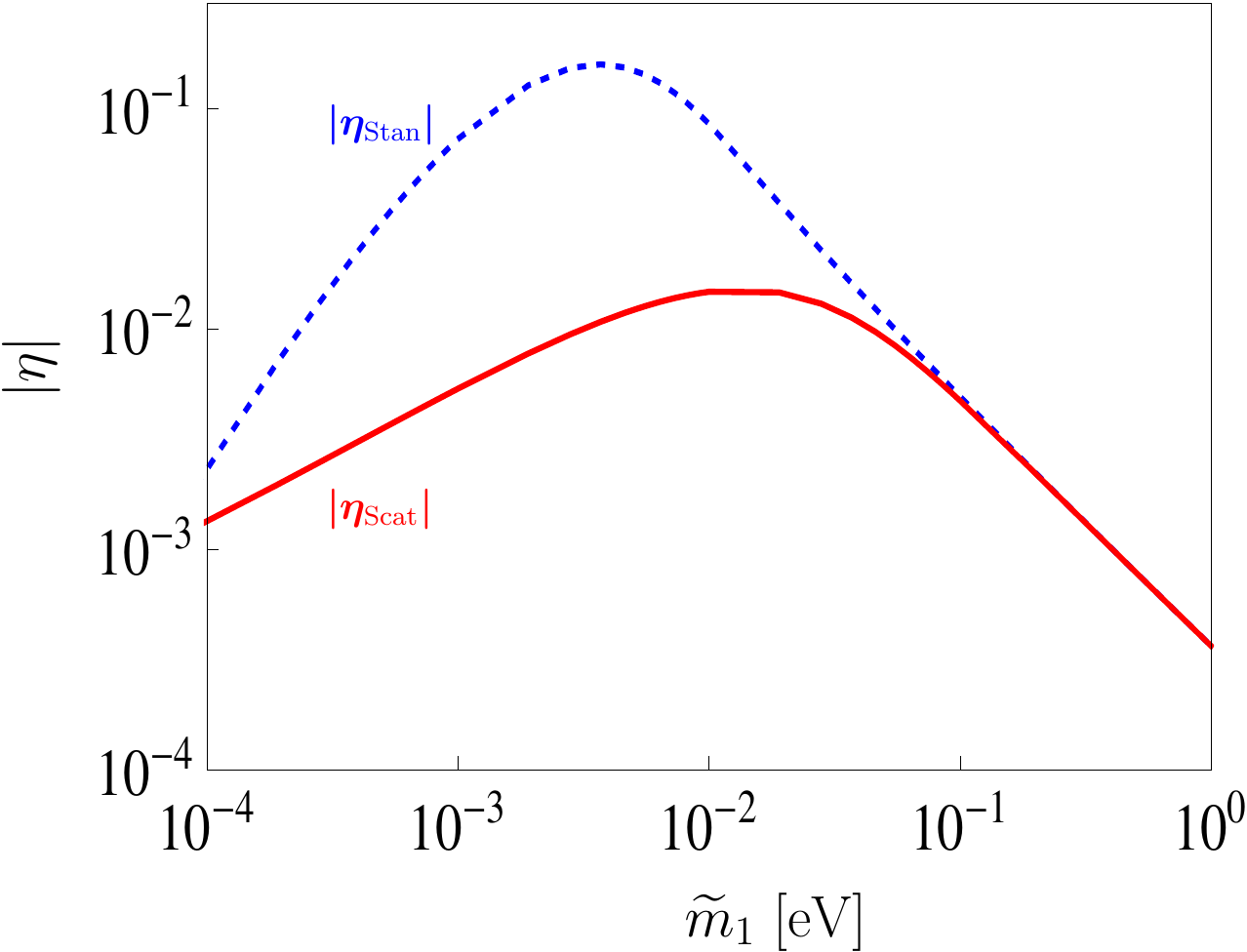}
  \caption{Efficiency versus $\widetilde m_1$ for the scattering
    parameters fixed as in Fig.~\ref{fig:reaction-densities}.  The red
    (solid) curve corresponds to the efficiency obtained when RH
    neutrino scatterings are included while the blue curve (dotted) to
    the efficiency in the standard case. }
  \label{fig:eff-vs-mtilde}
\end{figure}

The result in Fig. \ref{fig:eff-vs-z} exhibits an additional feature.
In the absence of the scalar-mediated RH neutrino scattering, i.e. if
$\gamma_S \ll \gamma_D$ (standard case) and in the weak washout
regime, say $\widetilde m_1=10^{-4}\,$~eV, and $Y_N^{\rm in}=0$, one finds that
the efficiency at freeze-out is determined by $\eta\sim \widetilde
m_1/m_\star$ (see left~panel in Fig.~\ref{fig:eff-vs-z}), as
expected~\cite{Giudice:2003jh}. However, one sees in the right panel
of Fig. \ref{fig:eff-vs-z} that the efficiency is almost maximal for
$\widetilde m_1=10^{-4}\,$~eV.
This enhancement can be traced back to the way in which the thermal
bath becomes populated with RH neutrinos at early epochs: regardless
of the strength of the scattering process, if at high temperatures
(small $z$'s) the condition $\gamma_S>\gamma_D$ is satisfied, then the
RH neutrino population becomes dictated by these processes, reaching
large values more rapidly than if determined by inverse decays. The
overall effect of the RH annihilation reactions can then be
``dissected'' as follows:
\begin{itemize}
\item {\it Enhancement of the efficiency}: if at early epochs the
  condition $\gamma_S>\gamma_D$ is satisfied and the scalar-induced RH
  neutrino scatterings are slow throughout the relevant temperature
  range, the efficiency is enhanced, readily reaching almost maximal
  values. This can lead to a dramatic enhancement in the weak washout
  regime.
\item {\it Suppression of the efficiency}: if at high temperatures the
  condition $\gamma_S>\gamma_D$ is satisfied, but during a certain
  period the RH neutrino scattering processes are fast, the efficiency
  can be strongly suppressed due to the RH neutrino thermalization
  induced by these reactions.
\end{itemize}
These effects are illustrated in Fig.~\ref{fig:eff-vs-mtilde}, for the
same parameter choice as in Fig. \ref{fig:reaction-densities}.
The case of {\it efficiency enhancement}
(Fig.~\ref{fig:eff-vs-mtilde}, left) can be understood from
Fig. \ref{fig:reaction-densities}, left, as follows.
Although $\gamma_S>\gamma_D$ holds at high temperatures even in the
strong washout regime, for $\widetilde m_1>10^{-3}$~eV the inverse
decay becomes dominant at lower temperatures.
In this case the efficiency matches the standard result, as seen in
the left panel of Fig.~\ref{fig:eff-vs-mtilde}.
In contrast, for parameters in the weak washout regime the RH neutrino
population is determined by the scalar-induced RH neutrino
scatterings, too slow to thermalize the RH neutrino distribution for a
long period, thus resulting in an striking enhancement of the
efficiency.
Turning to the case of {\it efficiency suppression}
(Fig.~\ref{fig:eff-vs-mtilde}, right) one finds that, due to fast
scatterings, RH neutrinos follow a thermal distribution down to
significantly low temperatures (large $z$). Thus, although the number
of RH neutrinos present in the heat bath is large, the overall
efficiency is far smaller as a result of RH neutrino thermalization
caused by the fast RH scalar-induced neutrino scatterings.
Note however that for large enough values of $\widetilde m_1$ the
generation of the $B-L$ asymmetry proceeds as in the standard
case. The reason is again that for values of $\widetilde m_1$ deep
inside the strong washout regime, the RH neutrino inverse decay is
still operative and dominant when the scattering interactions are
decoupled.

\subsection{Confronting observation}
\label{sec:confr-observ}

In the presence of rapid fermion-number violation due to
non-perturbative electroweak effects~\cite{kuzmin:1985mm}, the baryon
number of the Universe is obtained from the $B-L$ asymmetry of the
Universe generated as a result of the right-handed neutrino decays.
Using the standard result~\cite{harvey:1990qw}
\begin{equation}
Y_{\Delta_B}=(12/37)\,Y_{\Delta_{B-L}} \, ,
\end{equation}
 and the experimental value $Y_{\Delta_B}\subset [8.52,8.98]\times
 10^{-11}$~\cite{Hinshaw:2012aka,Ade:2013zuv} one can derive, from
 Eq.~(\ref{eq:BmL-formal-exp}), a lower limit for the efficiency in
 terms of the CP asymmetry factor $\epsilon_N$
\begin{equation}
  \label{eq:eff-lower-limit}
  \eta\gtrsim \frac{7\times 10^{-8}}{\epsilon_N}\ .
\end{equation}
One sees that, depending on the CP asymmetry parameter, largely
model-dependent, one has a minimum required value for the efficiency in
order to generate the correct baryon asymmetry.
For example, for the unrealistic case of maximal CP
asymmetry~\footnote{For strongly hierarchical RH neutrinos, such
  values are not possible without new contributions to the CP
  asymmetry~\cite{Davidson:2002qv}.}, $\epsilon_N=1$, any
parameter choice for which the efficiency factor drops below $\sim
7\times 10^{-8}$ will be unacceptable. Thus, for a given strength of
the scalar couplings one can determine the allowed $(\widetilde
m_1,M_{N_1})$ values for which the resulting $B-L$ is acceptable.

We have seen that the presence of the scalar-induced RH neutrino
scattering processes can have dramatic consequences in the weak
washout regime, either enhancing or suppressing the efficiency
depending on the scalar coupling parameter $\bar g$. The latter
suggests that there should exist parameter choices for which the
thermalization, induced by the annihilation process, may be still
consistent with the generation of an adequate baryon asymmetry.
For example for $M_{N_1}=10^{10}$~GeV, even maximizing the scattering
rate does not necessarily render the $B-L$ asymmetry below the
required value (Fig.~\ref{fig:eff-vs-mtilde}, right plot). In the
range where the RH neutrino scattering is effective $\widetilde
m_1\subset [10^{-4},10^{-1}]$~eV, the smallest value the efficiency has
is about $10^{-3}$, and therefore according to
Eq.~(\ref{eq:eff-lower-limit}) a CP asymmetry factor of order
$10^{-4}$ would suffice.
This exercise shows that in terms of the RH neutrino mass and
scattering cross section (see
Eqs. (\ref{eq:x-section-complete1})-(\ref{eq:x-section-complete3})),
one can identify, given certain value for the CP asymmetry, those
scenarios for which the efficiency depleting effect becomes
``harmless'' no matter the parameter choice, and those for which the
depleting effect render the $B-L$ asymmetry generation mechanism
ineffective.

\begin{figure}
  \centering
  \includegraphics[scale=0.65]{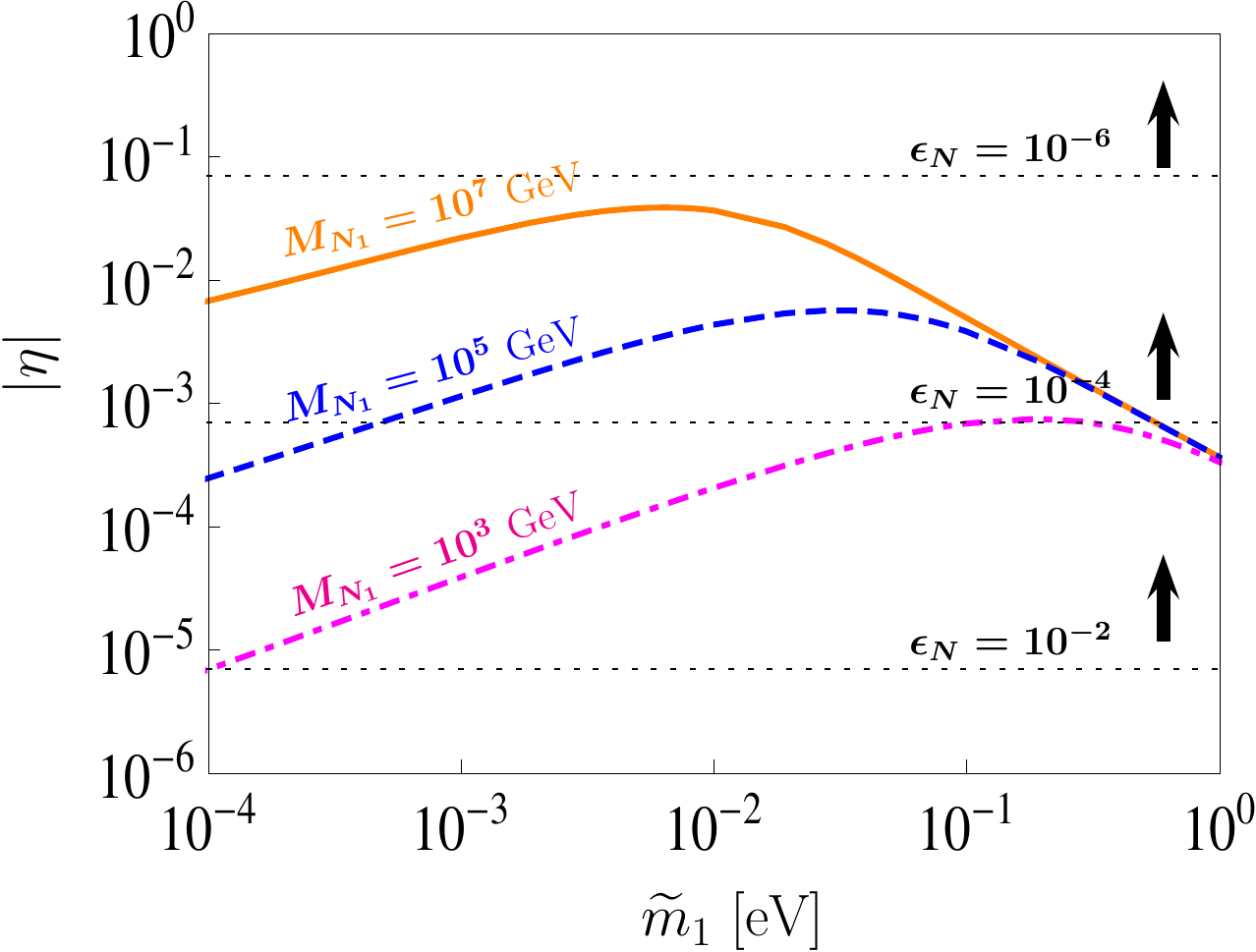}
  \hfill
  \includegraphics[scale=0.65]{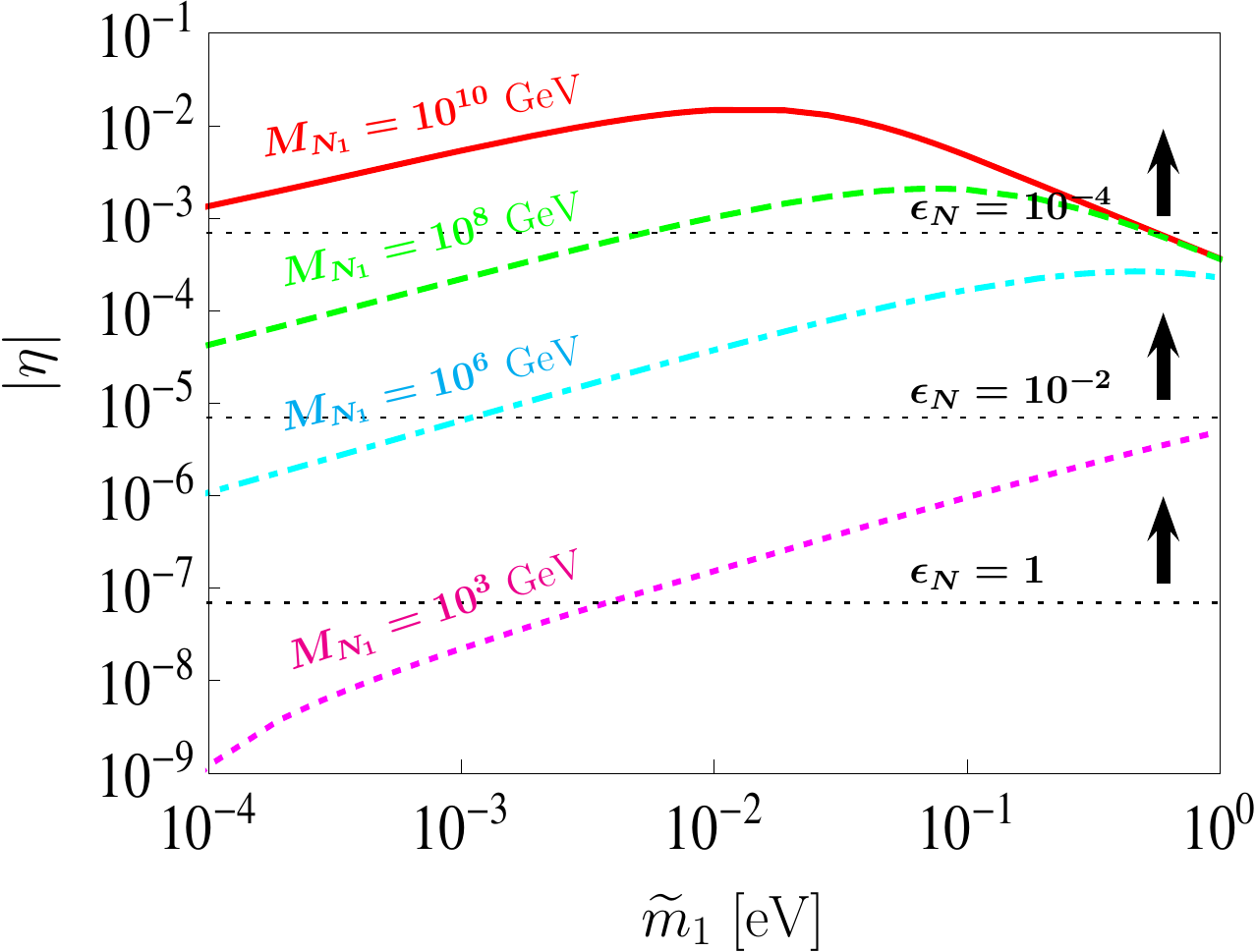}
  \caption{Efficiency as a function of $\widetilde m_1$ for several
    values of the RH neutrino mass and two different (representative)
    choices of the ``universal'' coupling $\bar g=10^{-1}$ left plot
    while $\bar g=1$ right plot. The horizontal lines indicate the
    minimum value that, for a given CP asymmetry, the efficiency
    parameter should have in order for the $B-L$ mechanism to
    successfully reproduce the measured $B$ asymmetry.}
  \label{fig:eff-vs-mtilde-diff-gbar}
\end{figure}

We illustrate quantitatively the requirements for a successful
leptogenesis scenario in Fig. \ref{fig:eff-vs-mtilde-diff-gbar}, which
shows the efficiency for two representative values of the
``universal'' coupling, $\bar g=10^{-1}$ for the left plot and $\bar g=1$ for
the right plot, and different RH neutrino masses. The horizontal
dotted lines indicate the minimum required efficiency values, given a
fixed CP asymmetry factor, see Eq. (\ref{eq:eff-lower-limit}). Points
in the efficiency curves lying below (above) those lines indicate
RH neutrino masses for which the leptogenesis mechanism fails (succeeds)
in accounting for the baryon asymmetry. From this one can draw several
generic conclusions: ($i$) if the CP asymmetry can be very large
($\epsilon_N\gtrsim 10^{-2}$), the scalar-induced RH neutrino
scattering does not place any significant constraint as seen in the
left plot; ($ii$) for the interesting range $\bar g\subset [10^{-1},1]$,
RH neutrinos with masses obeying $M_{N_1}>10^{8}$~GeV
($M_{N_1}>10^{11}$~GeV) can produce an adequate $B-L$ asymmetry
provided the CP asymmetry exceeds $10^{-6}$ and $\bar g=10^{-1}$
($\bar g=1$). In short, the scalar-induced RH neutrino scattering
constraints become most relevant in those scenarios where getting a
large CP asymmetry is not viable.

\subsection{Beyond the simplified scenario}
\label{sec:beyond-simpl-scen}

\begin{figure}[!t]
  \centering
  \includegraphics[scale=0.7]{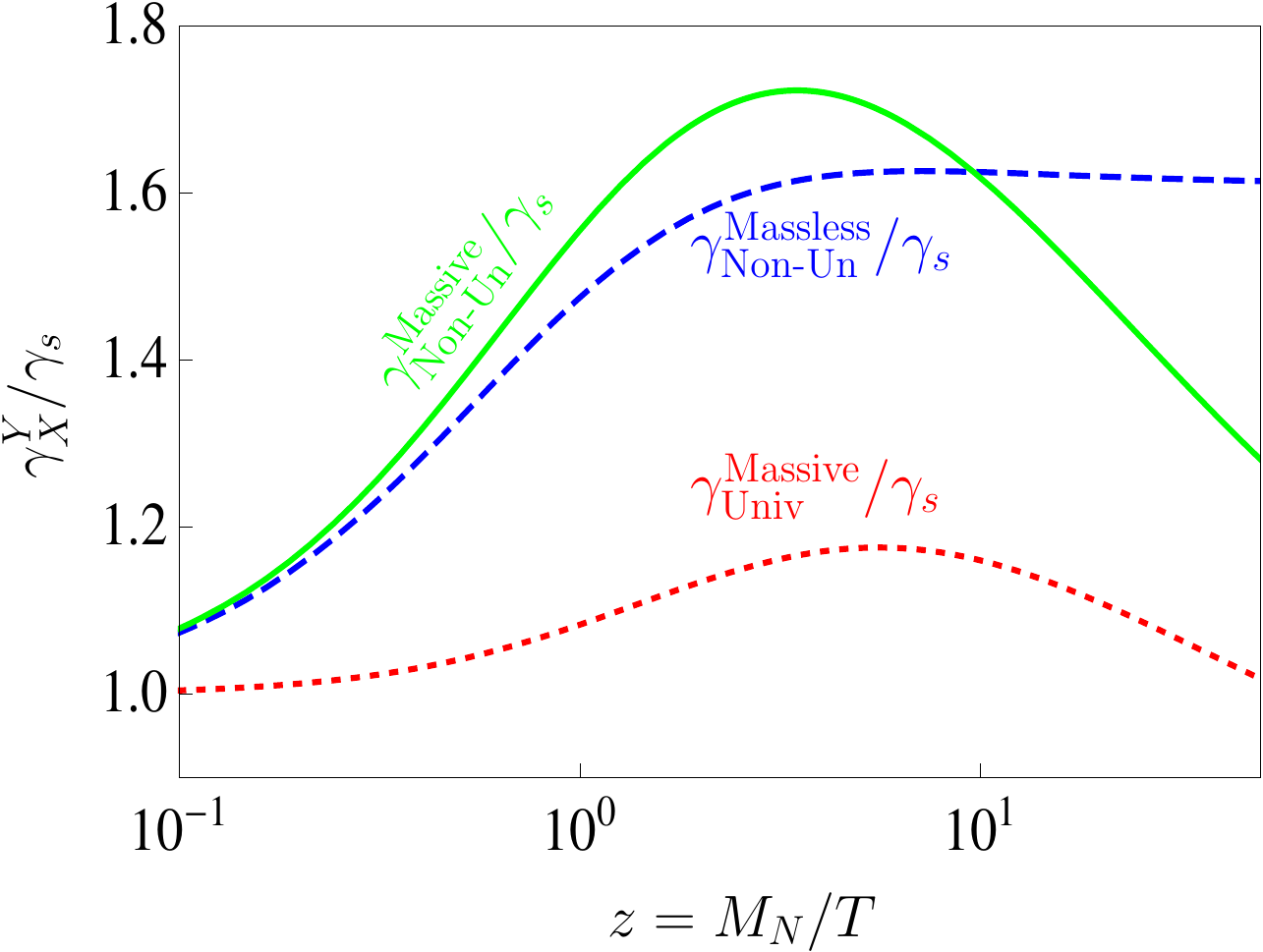}
  \caption{Ratios of right-handed neutrino scattering reaction
    densities as a function of $z=M_N/T$ for various scenarios. Solid
    (green) corresponds to massive scalars ($m_{S_1}=M_{N_1}$) and
    non-universal couplings; dashed (blue) to massless scalars and
    non-universal couplings; dotted (red) to massive scalars
    ($m_{S_1}=M_{N_1}$) and universal couplings. In all cases the
    normalization is given by the RH neutrino scattering reaction
    density calculated for massless scalars and universal
    couplings. See text for  details.}
  \label{fig:general-case}
\end{figure}
Finally, let us justify the usefulness of the simplified scenario with
massless spin zero states and universal dimensionless couplings. All
our previous numerical results have been obtained in this scenario and
thus the question arises as to whether they would be different, at
least qualitatively, in more general cases with massive scalars and
arbitrary coupling patterns. Fig. \ref{fig:general-case} answers this
question. It shows the annihilation reaction density as a function of
$z$ for several scenarios normalized to the one calculated in the
limit of massless scalars ($m_{S_a}\ll M_{N_1}$) and universal
couplings ($\bar g=1$). The cases correspond to: massive $S_1$, its
mass fixed as $m_{S_1}=M_{N_1}$, and non-universal couplings
$g_{NNa}=1$ and $g_{abc}=10^{-1}$ (solid/green curve); massless
scalars and non-universal couplings $g_{NNa}=1$ and $g_{abc}=10^{-1}$
(dashed/blue curve); massive $S_1$ and universal couplings (dotted/red
curve). This result demonstrates what we already anticipated in
Sec. \ref{sec:generalities}, namely deviations obtained as a result of
``non-universality'' and sizeable scalar masses are small, and so do
not change the general picture derived by using the simplest
scenario. This can be readily understood by inspecting the $N_1
N_1\leftrightarrow S_a\,S_a$ scattering cross section in
Eq. \eqref{eq:x-section-full}. The dominant contributions are the ones
given by the $t$- and $u$-channel diagrams, as can be seen by
comparing their contributions to the total cross section to the
$s$-channel contribution: $\sigma_{t,u}(s)/\sigma_s(s)\sim
2x_N+4\log(x_N)\gg 1$, where we have assumed universal couplings. The
only exception is found for $g_{NNa} \ll g_{abc}$, but in that case
the RH neutrino scattering process plays no role in the generation of
the $B-L$ asymmetry. Since the $t$- and $u$-channel diagrams only
depend on the $g_{NNa}$ couplings and are mediated by RH neutrino
exchange, the purely scalar couplings and the masses of the spin zero
states play a secondary role in the determination of the annihilation
reaction density, explaining why the simplified scenario gives a good
description.

\section{The case of spontaneous lepton number breaking}
\label{sec:spontaneous-lepton-number-breaking}

Let us now apply the results derived in the previous sections to the
simplest seesaw majoron model discussed in
Sec.~\ref{sec:intro}~\cite{chikashige:1981ui,Schechter:1981cv}.
Type-I seesaw majoron leptogenesis has already been partially analyzed
in Refs.~\cite{Gu:2009hn,Pilaftsis:2008qt}. We start the analysis by
discussing the main features of this scheme.


\subsection{Scalar sector}
\label{sec:scalar-sector}

The extended scalar sector of the model consists of the Higgs doublet,
$H$, and a new \SM singlet complex scalar field, $\sigma$, carrying
lepton number charge $L=-2$. For simplicity we assume CP invariance in
the scalar potential. After the electroweak gauge symmetry and lepton
number get broken, one gets the following mass terms for the neutral
scalars ($\rho$) and pseudoscalars ($\kappa$)
\begin{equation}
  \label{eq:Vquad}
  V_{\text{quadratic}} \supset 
  \frac{1}{2} \rho_a \left(M_S^2\right)_{ab} \rho_b 
  + \frac{1}{2} \kappa_a \left(M_P^2\right)_{ab} \kappa_b \ ,
\end{equation}
where a sum over the scalar indices $a,b = 1,2$ has been left
implicit. In the basis $\rho = \text{Re} \left( \sigma , \phi^0
\right)^T$ and $\kappa = \text{Im} \left( \sigma , \phi^0 \right)^T$
the mass matrices $M_S^2$ and $M_P^2$ are given by
\begin{align}
  \label{eq:scalar-mass-matrices}
  M_S^2 &=
  \left( \begin{array}{cc}
      \frac{1}{4} \left(6 \lambda_\sigma u^2 - 2 m_\sigma^2 + \delta v^2 \right) 
      & \frac{1}{2} \delta v u \\
      \frac{1}{2} \delta v u 
      & \frac{1}{4} \left( \delta u^2 - 2 m_H^2 + 6 \lambda_H v^2 \right)
    \end{array} \right)\ ,\\
  M_P^2 &= 
  \left( \begin{array}{cc}
      \frac{1}{4} \left(2 \lambda_\sigma u^2 - 2 m_\sigma^2 + \delta v^2 \right) 
      & 0 \\
      0 
      & \frac{1}{4} \left( \delta u^2 - 2 m_H^2 + 2 \lambda_H v^2 \right)
    \end{array} \right) \, .
\end{align}
The neutral scalar mass matrix $M_S^2$ becomes diagonal by going to
the mass basis $S = \left( S_1 , S_2 \right)^T = R_S^\dagger \rho$,
where $R_S$ is a unitary matrix such that $R_S M_S^2 R_S^\dagger =
\text{diag}(m_{S_1}^2 , m_{S_2}^2)$. After applying the minimization
conditions for the scalar potential in Eq. \eqref{eq:Vscalar}, the
resulting mass eigenvalues and unitary matrix $R_S$ are found to be
\begin{eqnarray}
  m_{S_{1,2}}^2 &=&
  \frac{1}{2} 
  \left( \lambda_\sigma u^2 \pm \tilde u^2 + \lambda_H v^2 \right)\ ,
  \\
  R_S &=& 
  \left( \begin{array}{cc}
      -\frac{\lambda_H v^2-\tilde u^2-\lambda_\sigma u^2}
      {\sqrt{\delta^2 u^2 v^2 
          + \left( \lambda_H v^2 
            - \tilde u^2 - \lambda_\sigma u^2 \right)^2}} 
      & \frac{\delta u v}
      {\sqrt{\delta^2 u^2 v^2 + \left( \lambda_H v^2 
            - \tilde u^2 
            - \lambda_\sigma u^2 \right)^2}}
      \\
      \frac{-\lambda_H v^2 - \tilde u^2 + \lambda_\sigma u^2}
      {\sqrt{\delta^2 u^2 v^2 + \left( -\lambda_H v^2 - \tilde u^2 
            + \lambda_\sigma u^2 \right)^2}} 
      & \frac{\delta u v}{\sqrt{\delta^2 u^2 v^2 + 
          \left( -\lambda_H v^2 - \tilde u^2 
            + \lambda_\sigma u^2 \right)^2}}
    \end{array} \right) \, ,
\end{eqnarray}
where we have defined
\begin{equation}
\tilde u^2 = 
\sqrt{\lambda_\sigma^2 u^4 + u^2 v^2 
  (\delta^2-2 \lambda_H \lambda_\sigma)
  +\lambda_H^2 v^4} \, .
\end{equation}
In the limit $\delta \ll 1$ (or, similarly, for $v \ll u$) the mixing
between the two scalar states becomes negligible. In this scenario
$S_1$ is mainly singlet, with a squared mass $m_{S_1}^2 \simeq
\lambda_\sigma u^2$. The other state, with a squared mass $m_{S_2}^2
\equiv m_h^2 \simeq \lambda_H v^2$, is identified as the standard
model Higgs boson. On the other hand, the pseudoscalar mass matrix
$M_P^2$ contains two vanishing eigenvalues. After minimizing the
scalar potential it simplifies to a $2 \times 2$ matrix with vanishing
entries. As expected, the spontaneous breaking of lepton number
implies the existence of a massless physical Nambu-Goldstone boson,
the majoron $J$, with a pure singlet nature. The other massless
pseudoscalar state corresponds to the would-be Goldstone boson that
becomes the longitudinal component of the $Z$ boson. We can then make
the following identification $P = \left( P_1 , P_2 \right)^T \equiv \left(
  J , G^0 \right)^T$ and $S_3 \equiv P_1$, $S_3$ being the generic
pseudoscalar field introduced in Sec. \ref{sec:generalities}.

In order to identify the $g_{NNa}$ and $g_{abc}$ couplings used in
Sec.~\ref{sec:BmL} in terms of the ``fundamental'' parameters of the
model, one must find the corresponding expressions for these
couplings, namely~\footnote{In this section we will use a simplified
  notation in which $h$ actually refers to $h_{11}$, the element of
  the $h$ Yukawa matrix that involves two $N_1$'s.}
\begin{align}
  \label{eq:coupTypeI}
  g_{N N 1} &= - \frac{h}{\sqrt{2}} R_S^{11}\ ,
  \\
  g_{N N 2} &= - \frac{h}{\sqrt{2}} R_S^{12}\ ,
  \\
  \mu_{1 1 1} &= \frac{1}{2} 
  \left[ 2 u \lambda_\sigma \left(R_S^{11}\right)^3
    + \delta u R_S^{11} \left(R_S^{21}\right)^2 
    + \delta \left(R_S^{11}\right)^2 R_S^{21} v + 
    2 \lambda_H \left(R_S^{21}\right)^3 v \right]\ ,
  \\
  \mu_{1 1 2} &= \frac{1}{2} 
  \left[u \left( 6 \lambda_\sigma \left(R_S^{11}\right)^2 R_S^{12}
      + 2 \delta R_S^{11} R_S^{21} R_S^{22}+ \delta R_S^{12} 
      \left(R_S^{21}\right)^2 \right) \right.
  \nonumber \\
  & \left. + v \left( \delta \left(R_S^{11}\right)^2 R_S^{22}
      +2 \delta R_S^{11} R_S^{12} R_S^{21}+6 \lambda_H 
      \left(R_S^{21}\right)^2 R_S^{22} \right) \right] \ ,
  \\
  \mu_{1 2 2} &= 
  \frac{1}{2} \left[u \left( 6 \lambda_\sigma R_S^{11} 
      \left(R_S^{12}\right)^2 + \delta R_S^{11} 
      \left(R_S^{22}\right)^2+ 2 \delta R_S^{12} R_S^{21} R_S^{22} \right)
  \right.
  \nonumber \\
  & \left. + 
    v \left( 2 \delta R_S^{11} R_S^{12} R_S^{22}+ \delta 
      \left(R_S^{12}\right)^2 R_S^{21}+6 \lambda_H  R_S^{21} 
      \left(R_S^{22}\right)^2 \right) \right]\ ,
  \\
  \mu_{2 2 2} &= \frac{1}{2} 
  \left[ 2 u \lambda_\sigma \left(R_S^{12}\right)^3
    + \delta u R_S^{12} \left(R_S^{22}\right)^2 
    + \delta \left(R_S^{12}\right)^2 R_S^{22} v
    +2 \lambda_H \left(R_S^{22}\right)^3 v \right]\ , 
  \\
  \mu_{1 3 3} &= \lambda_\sigma u R_S^{11} 
  + \frac{1}{2} \delta v R_S^{21}\ ,
  \\
  \label{eq:mu233}
  \mu_{2 3 3} &= \lambda_\sigma u R_S^{12} 
  + \frac{1}{2} \delta v R_S^{22}\ .
\end{align}
The above relations simplify in the limit in which the
singlet-doublet mixing is negligible (given by $\delta \ll 1$ or $v
\ll u$). In this case $R_S$ is the $2 \times 2$ identity matrix and
one finds
\begin{alignat}{2}
  \label{eq:coupTypeILimit}
  g_{N N 1} &= - \frac{h}{\sqrt{2}}\ ,
  &\qquad\qquad
  g_{N N 2}&= 0\ ,\\
  \mu_{1 1 1} &= \mu_{1 3 3} = u \lambda_\sigma \ ,
  &\qquad\quad
  \mu_{1 1 2}&= \mu_{2 3 3} = \frac{1}{2} v \delta \ ,\\
  \mu_{1 2 2} &=
  \frac{1}{2} u \delta \ ,
  &\qquad\qquad
  \mu_{2 2 2}&= \lambda_H v \ .
\end{alignat}
The explicit efficiency calculation presented in Sec.~\ref{sec:BmL}
reveals that the most important parameters can be chosen as the
right-handed neutrino mass $M_{N_1}$ and its coupling to the scalars
$g_{NNa}$.


\subsection{The $B-L$ asymmetry in the presence of majorons}

\begin{figure}
  \centering
  \includegraphics[scale=0.65]{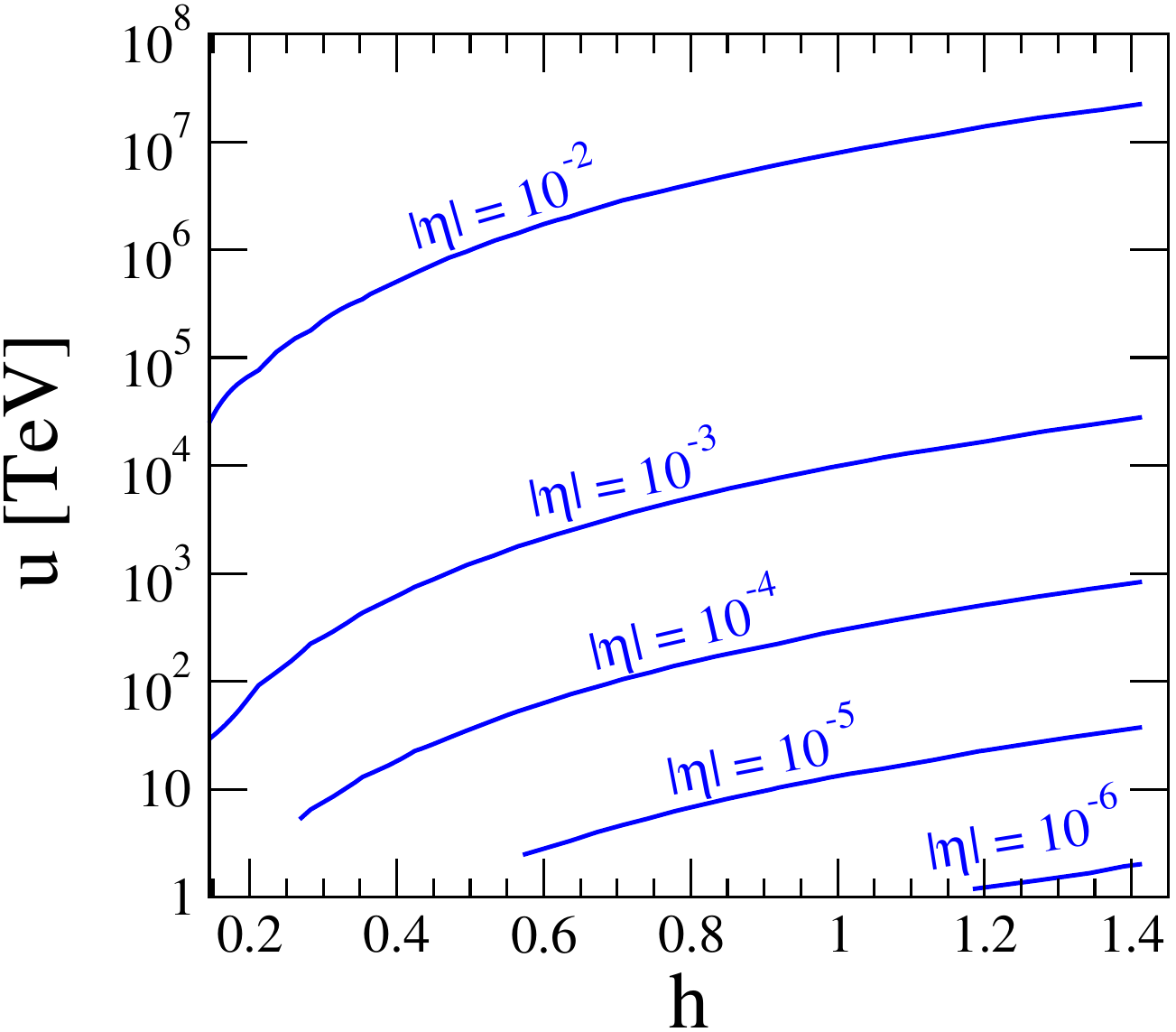}
  \hfill
  \includegraphics[scale=0.65]{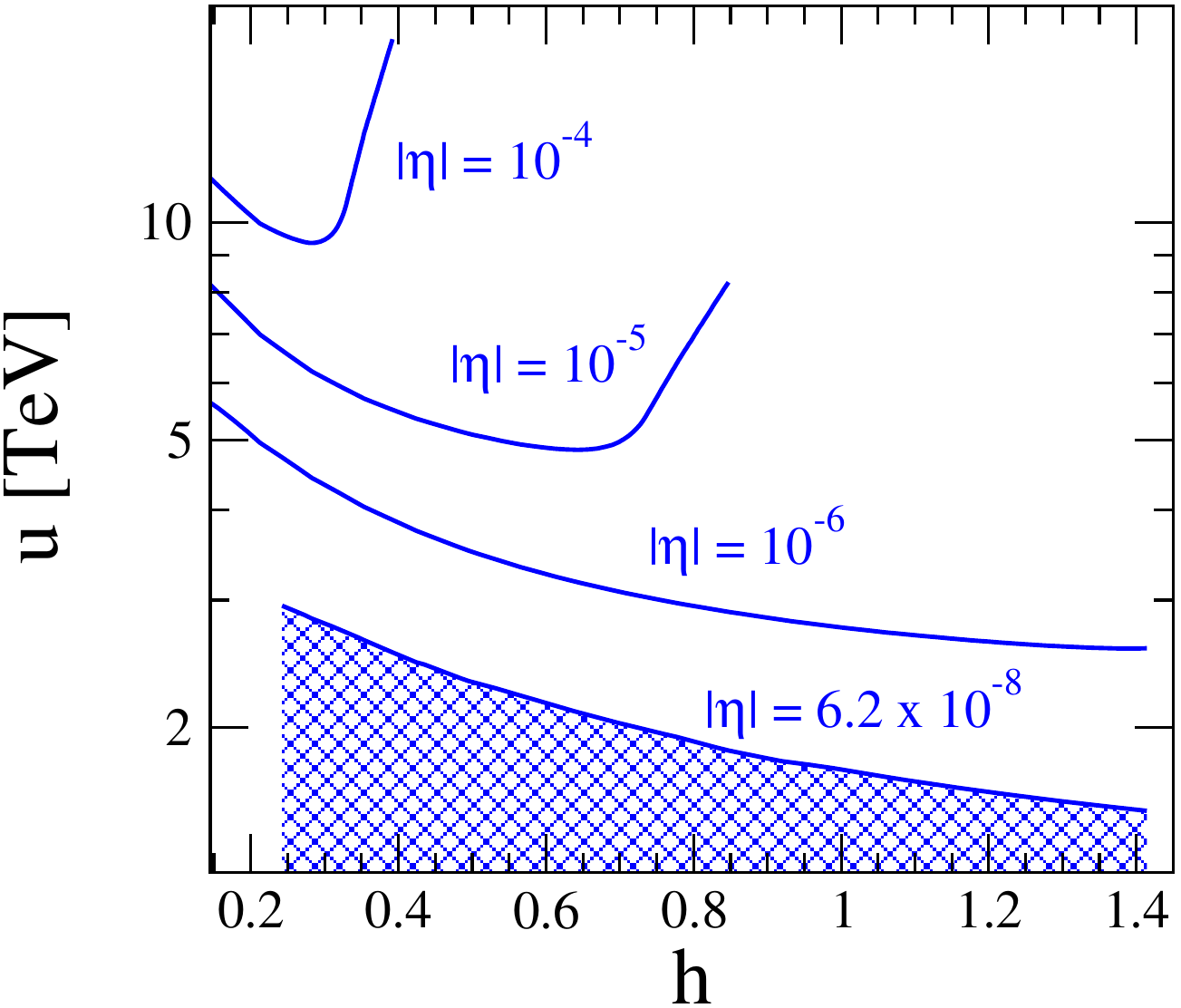}
  \caption{Efficiency contours in the $h$-$u$ plane, for a fixed value
    of $\widetilde m_1 = 10^{-3}$ eV. In the left panel we show our
    results for $M_N > 5$ TeV, while on the right panel we focus on
    the region with lower right-handed neutrino masses. See text for
    details.}
  \label{fig:contours}
\end{figure}

We now turn to the interpretation of our results in terms of the
relevant majoron model parameters, chosen as the scale of spontaneous
\lnv $u$ and the Yukawa coupling $h$. In the left panel in
Fig. \ref{fig:contours} we show our results for the efficiency in the
$h$-$u$ plane, obtained for a fixed value of $\widetilde m_1 =
10^{-3}$ eV, for RH neutrino masses in the range $[5 \cdot 10^3 ,
10^{10}]$ GeV. 
On the other hand in the right panel we have focused on RH neutrino
masses in the range $M_N < 5$ TeV. These figures can be easily
interpreted using Eq. \eqref{eq:eff-lower-limit}. Each efficiency
value can be translated into a necessary value for the CP asymmetry,
$\epsilon_N$, which allows for an adequate value for the baryon
asymmetry of the Universe. For example, the isocurve for which $\eta =
10^{-6}$ requires $\epsilon_N \sim 10^{-3}$. The contour with $\eta =
6.2 \times 10^{-8}$ shown on the right panel is the lower limit of the
viable parameter space region, since it requires a maximal CP
asymmetry obeying $\epsilon_N \sim 1$. Points below that contour
(hatched region) are thus excluded by the measured baryon asymmetry
combined with the $B-L$ yield derived from leptogenesis. One sees the
trouble in reconciling a sizeable baryon asymmetry generated through
leptogenesis within a minimal seesaw majoron scheme.
Note that for low right-handed neutrino masses $M_N < 5$ TeV the
$2\leftrightarrow 2$ RH neutrino annihilation processes are fast down
to low temperatures (large $z$). In this case the RH neutrinos are
almost thermalized by these scatterings and hence unable to generate a
sufficiently large $B-L$ asymmetry before sphaleron decoupling which,
for a Higgs mass of about $125\,$~GeV, takes place at about $T\sim
135\,$~GeV \cite{Strumia:2008cf,donofrio:2014kta}. This means that the
electroweak sphaleron processes freeze out before converting the $B-L$
asymmetry into an adequate baryon asymmetry.
This effect should be taken into account whenever dealing with a state
whose mass is near the electroweak scale and subject to fast
reactions, as it might be the case in low-scale leptogenesis
models. Implementation of these scenarios require overcoming the upper
limit of the CP asymmetry~\cite{Davidson:2002qv}, something which can
be in principle done by resorting to extended models with additional
contributions to the CP asymmetry (not subject to constraints from
neutrino masses), see for
example~\cite{AristizabalSierra:2007ur,Fong:2013gaa,Sierra:2013kba},
or by assuring a quasi-degenerate RH neutrino mass spectrum leading to
resonant effects~\cite{Flanz:1996fb,
  Covi:1996fm,Pilaftsis:1997jf}. These scenarios are beyond the scope
of this paper, and so we will not add further details on these
specific realizations.

\section{Conclusions}
\label{sec:conclusions}

Seesaw schemes with spontaneous \lnv generate neutrino mass while
potentially addressing other cosmological issues such as the origin of
dark matter and inflation. They can also account for the cosmic baryon
asymmetry via leptogenesis. These models all require extended Higgs
sectors in order to realize an adequate symmetry breaking pattern
involving lepton number as well as electroweak breaking. For the
simplest case of the standard \21 gauge structure and ungauged lepton
number, a key feature is the presence of a (pseudo) Nambu-Goldstone
boson, the majoron $J$, associated to the spontaneous breaking of
lepton number.
Inspired by the attractiveness of such schemes, we have analyzed the
effects that the novel right-handed neutrino scalar-mediated
annihilation processes ($NN\leftrightarrow S_a S_a$) can have in the
generation of the $B-L$ asymmetry.
Even in simple realizations, these processes involve many free
parameters. However, we have identified the two new parameters which
best describe the implications of such scatterings in the generation
of the $B-L$ asymmetry.
Within our simple description we have shown that scattering processes
can change the conventional picture. First note that the heat bath
will be populated with the intervening scalars, either produced by
reheating or via their couplings with the standard model Higgs. In
this case the RH neutrino density in the thermal bath will be
determined by the novel scattering reactions rather than by inverse
decays.
In the parameter regions where these reactions are slow in the
relevant temperature range, their presence will tend to enhance the
$B-L$ yield, provided the Dirac neutrino Yukawa couplings lie in the
weak washout regime. A direct consequence of this is that of the $B-L$
asymmetry will no longer depend upon initial conditions, something
inherent to the ``standard'' picture in the weak washout regime.  In
contrast, if the new scatterings are fast the RH neutrino distribution
will closely follow a thermal distribution. Although such effect will
tend to suppress the $B-L$ yield, it need not invalidate the $B-L$
production mechanism, provided the CP-asymmetry is large enough. In
any case, a proper treatment of the $B-L$ asymmetry generation
mechanism within these schemes requires the inclusion of RH neutrino
annihilations, as long as the relevant couplings are sizeable.

Before closing let us mention that the presence of Goldstone bosons,
such as the majoron, if associated to low-scale symmetry breaking, may
give rise to other non-standard signatures. For example, it can induce
a large standard model Higgs invisible decay
mode~\cite{joshipura:1992hp}. The restrictions on the parameter space
are determined by the upper limit on the invisible decay width of the
Higgs boson.  The current upper bound given by the ATLAS
collaboration~\cite{Aad:2014iia} for the invisible decay mode
branching ratio is not too stringent, allowing for invisible modes
ranging from 10\% up to 75\%, and thus leaving large fractions of
parameter space still open.  In the next LHC run a better sensitivity
to this mode will be possible. On the other hand, these schemes may
also lead to exotic lepton flavor violating decays with majoron
emission, such as $\mu \to e J$ or $\mu \to e J
\gamma$~\cite{Romao:1991tp,Hirsch:2009ee,GarciaiTormo:2011et}.
In any case all these possibilities require low \lnv scale, which do
not fit within the minimal dynamical leptogenesis schemes considered
above. Extended seesaw leptogenesis schemes with right-handed neutrino
scatterings lie outside the scope of this paper.

\section{Acknowledgments}
DAS and AV would like to thank Juan Racker for useful comments. This
work was supported by MINECO grants FPA2011-22975, MULTIDARK
Consolider CSD2009-00064. DAS is supported by a ``Charg\'e de
Recherches'' contract funded by the Belgian FNRS agency. AV is
partially supported by the EXPL/FIS-NUC/0460/2013 project financed by
the Portuguese FCT.

\bibliographystyle{h-physrev4.bst} 

\begin{thebibliography}{10}

\bibitem{Tortola:2012te}
D.~Forero, M.~Tortola and J.~W.~F. Valle,
\newblock Phys.Rev. {\bf D86}, 073012 (2012), [arXiv:1205.4018];
for review with a comprehensive set of references see
M.~Maltoni, T.~Schwetz, M.~Tortola and J.~Valle,
\newblock New J.Phys. {\bf 6}, 122 (2004), [hep-ph/0405172].

\bibitem{Hinshaw:2012aka}
WMAP, G.~Hinshaw {\em et~al.},
\newblock Astrophys.J.Suppl. {\bf 208}, 19 (2013), [1212.5226].

\bibitem{Ade:2013zuv}
Planck Collaboration, P.~Ade {\em et~al.},
\newblock 1303.5076.

\bibitem{minkowski:1977sc}
P.~Minkowski,
\newblock Phys. Lett. {\bf B67}, 421 (1977).

\bibitem{gell-mann:1980vs}
M.~Gell-Mann, P.~Ramond and R.~Slansky,
\newblock (1979),
\newblock Print-80-0576 (CERN).

\bibitem{yanagida:1979}
T.~Yanagida,
\newblock (KEK lectures, 1979),
\newblock ed. O. Sawada and A. Sugamoto (KEK, 1979).

\bibitem{Lazarides:1980nt}
G.~Lazarides, Q.~Shafi and C.~Wetterich,
\newblock Nucl. Phys. {\bf B181}, 287 (1981).

\bibitem{mohapatra:1980ia}
R.~N. Mohapatra and G.~Senjanovic,
\newblock Phys. Rev. Lett. {\bf 44}, 91 (1980).

\bibitem{Schechter:1980gr}
J.~Schechter and J.~W.~F. Valle,
\newblock Phys. Rev. {\bf D22}, 2227 (1980).

\bibitem{mohapatra:1986bd}
R.~N. Mohapatra and J.~W.~F. Valle,
\newblock Phys. Rev. {\bf D34}, 1642 (1986).

\bibitem{akhmedov:1995ip}
E.~Akhmedov {\em et~al.},
\newblock Phys. Lett. {\bf B368}, 270 (1996), [hep-ph/9507275].

\bibitem{akhmedov:1995vm}
E.~Akhmedov {\em et~al.},
\newblock Phys. Rev. {\bf D53}, 2752 (1996), [hep-ph/9509255].

\bibitem{Malinsky:2005bi}
M.~Malinsky, J.~C. Romao and J.~W.~F. Valle,
\newblock Phys. Rev. Lett. {\bf 95}, 161801 (2005).

\bibitem{chikashige:1981ui}
Y.~Chikashige, R.~N. Mohapatra and R.~D. Peccei,
\newblock Phys. Lett. {\bf B98}, 265 (1981).

\bibitem{Schechter:1981cv}
J.~Schechter and J.~W.~F. Valle,
\newblock Phys. Rev. {\bf D25}, 774 (1982).

\bibitem{gelmini:1984ea}
G.~B. Gelmini and J.~W.~F. Valle,
\newblock Phys. Lett. {\bf B142}, 181 (1984).

\bibitem{gonzalezgarcia:1989rw}
M.~C. Gonzalez-Garcia and J.~W.~F. Valle,
\newblock Phys. Lett. {\bf B216}, 360 (1989).

\bibitem{kallosh:1995hi}
R.~Kallosh, A.~D. Linde, D.~A. Linde and L.~Susskind,
\newblock Phys. Rev. {\bf D52}, 912 (1995), [hep-th/9502069].

\bibitem{berezinsky:1993fm}
V.~Berezinsky and J.~W.~F. Valle,
\newblock Phys. Lett. {\bf B318}, 360 (1993), [hep-ph/9309214].

\bibitem{Lattanzi:2007ux}
M.~Lattanzi and J.~W.~F. Valle,
\newblock Phys. Rev. Lett. {\bf 99}, 121301 (2007), [arXiv:0705.2406
  [astro-ph]].

\bibitem{Lattanzi:2013uza}
M.~Lattanzi {\em et~al.},
\newblock Phys.Rev. {\bf D88}, 063528 (2013), [1303.4685].

\bibitem{Queiroz:2014yna}
F.~S. Queiroz and K.~Sinha,
\newblock 1404.1400.

\bibitem{Boucenna:2014uma}
S.~M. Boucenna, S.~Morisi, Q.~Shafi and J.~W.~F. Valle,
\newblock 1404.3198.

\bibitem{Higaki:2014dwa}
T.~Higaki, R.~Kitano and R.~Sato,
\newblock 1405.0013.

\bibitem{Ade:2014xna}
BICEP2 Collaboration, P.~Ade {\em et~al.},
\newblock 1403.3985.

\bibitem{fukugita:1986hr}
M.~Fukugita and T.~Yanagida,
\newblock Phys. Lett. {\bf B174}, 45 (1986).

\bibitem{davidson:2008bu}
S.~Davidson, E.~Nardi and Y.~Nir,
\newblock Phys.Rept. {\bf 466}, 105 (2008), [0802.2962].

\bibitem{Fong:2013wr}
C.~S. Fong, E.~Nardi and A.~Riotto,
\newblock Adv.High Energy Phys. {\bf 2012}, 158303 (2012), [1301.3062].

\bibitem{kuzmin:1985mm}
V.~A. Kuzmin, V.~A. Rubakov and M.~E. Shaposhnikov,
\newblock Phys. Lett. {\bf B155}, 36 (1985).

\bibitem{valle:1987sq}
J.~W.~F. Valle,
\newblock Phys. Lett. {\bf B196}, 157 (1987).

\bibitem{Pati:1974yy}
J.~C. Pati and A.~Salam,
\newblock Phys. Rev. {\bf D10}, 275 (1974).

\bibitem{Chung:1998rq}
D.~J. Chung, E.~W. Kolb and A.~Riotto,
\newblock Phys.Rev. {\bf D60}, 063504 (1999), [hep-ph/9809453].

\bibitem{Hambye:2005tk}
T.~Hambye, M.~Raidal and A.~Strumia,
\newblock Phys. Lett. {\bf B632}, 667 (2006), [hep-ph/0510008].

\bibitem{Hambye:2012fh}
T.~Hambye,
\newblock New J.Phys. {\bf 14}, 125014 (2012), [1212.2888].

\bibitem{AristizabalSierra:2010mv}
D.~Aristizabal~Sierra, J.~F. Kamenik and M.~Nemevsek,
\newblock JHEP {\bf 1010}, 036 (2010), [1007.1907].

\bibitem{Sierra:2014tqa}
D.~Aristizabal~Sierra, M.~Dhen and T.~Hambye,
\newblock 1401.4347.

\bibitem{Gu:2009hn}
P.-H. Gu and U.~Sarkar,
\newblock Eur.Phys.J. {\bf C71}, 1560 (2011), [0909.5468].

\bibitem{Davidson:2002qv}
S.~Davidson and A.~Ibarra,
\newblock Phys. Lett. {\bf B535}, 25 (2002), [hep-ph/0202239].

\bibitem{Hambye:2003rt}
T.~Hambye, Y.~Lin, A.~Notari, M.~Papucci and A.~Strumia,
\newblock Nucl.Phys. {\bf B695}, 169 (2004), [hep-ph/0312203].

\bibitem{AristizabalSierra:2011ab}
D.~Aristizabal~Sierra, F.~Bazzocchi and I.~de~Medeiros~Varzielas,
\newblock Nucl.Phys. {\bf B858}, 196 (2012), [1112.1843].

\bibitem{Cosme:2004xs}
N.~Cosme,
\newblock JHEP {\bf 0408} 027 (2004), [hep-ph/0403209].

\bibitem{Frere:2008ct}
J.-M. Frere, T.~Hambye and G.~Vertongen,
\newblock JHEP {\bf 0901}, 051 (2009), [0806.0841].

\bibitem{Blanchet:2010kw} 
  S.~Blanchet, P.~S.~B.~Dev and R.~N.~Mohapatra,
  Phys.\ Rev.\ D {\bf 82}, 115025 (2010)
  [arXiv:1010.1471 [hep-ph]].

\bibitem{Plumacher:1996kc}
M.~Plumacher,
\newblock Z.\ Phys.\ C {\bf 74} 549 (1997), [hep-ph/9604229].

\bibitem{Racker:2008hp}
J.~Racker and E.~Roulet,
\newblock JHEP {\bf 0903} 065 (2009), [0812.4285].

\bibitem{Blanchet:2009bu} 
  S.~Blanchet, Z.~Chacko, S.~S.~Granor and R.~N.~Mohapatra,
  Phys.\ Rev.\ D {\bf 82}, 076008 (2010)
  [arXiv:0904.2174 [hep-ph]].

\bibitem{Buchmuller:2012wn} 
  W.~Buchmuller, V.~Domcke and K.~Schmitz,
  Nucl.\ Phys.\ B {\bf 862}, 587 (2012)
  [arXiv:1202.6679 [hep-ph]].

\bibitem{Schechter:1981hw}
J.~Schechter and J.~W.~F. Valle,
\newblock Phys. Rev. {\bf D24}, 1883 (1981),
\newblock Err. D25, 283 (1982).

\bibitem{wolfenstein:1981rk}
L.~Wolfenstein,
\newblock Phys. Lett. {\bf B107}, 77 (1981).

\bibitem{Giudice:2003jh}
G.~F. Giudice, A.~Notari, M.~Raidal, A.~Riotto and A.~Strumia,
\newblock Nucl. Phys. {\bf B685}, 89 (2004), [hep-ph/0310123].

\bibitem{Nardi:2007jp}
E.~Nardi, J.~Racker and E.~Roulet,
\newblock JHEP {\bf 0709}, 090 (2007), [0707.0378].

\bibitem{Barbieri:1999ma}
R.~Barbieri, P.~Creminelli, A.~Strumia and N.~Tetradis,
\newblock Nucl. Phys. {\bf B575}, 61 (2000), [hep-ph/9911315].

\bibitem{harvey:1990qw}
J.~A. Harvey and M.~S. Turner,
\newblock Phys. Rev. {\bf D42}, 3344 (1990).

\bibitem{Pilaftsis:2008qt}
A.~Pilaftsis,
\newblock Phys.Rev. {\bf D78}, 013008 (2008), [0805.1677].

\bibitem{Strumia:2008cf}
A.~Strumia,
\newblock Nucl.Phys. {\bf B809}, 308 (2009), [0806.1630].

\bibitem{donofrio:2014kta}
M.~D'Onofrio, K.~Rummukainen and A.~Tranberg,
\newblock 1404.3565.

\bibitem{AristizabalSierra:2007ur}
D.~Aristizabal~Sierra, M.~Losada and E.~Nardi,
\newblock Phys.Lett. {\bf B659}, 328 (2008), [0705.1489].

\bibitem{Fong:2013gaa}
C.~S. Fong, M.~Gonzalez-Garcia, E.~Nardi and E.~Peinado,
\newblock JHEP {\bf 1308}, 104 (2013), [1305.6312].

\bibitem{Sierra:2013kba}
D.~Aristizabal~Sierra, C.~S. Fong, E.~Nardi and E.~Peinado,
\newblock JCAP {\bf 1402}, 013 (2014), [1309.4770].

\bibitem{Flanz:1996fb}
M.~Flanz, E.~A. Paschos, U.~Sarkar and J.~Weiss,
\newblock Phys.Lett. {\bf B389}, 693 (1996), [hep-ph/9607310].

\bibitem{Covi:1996fm}
L.~Covi and E.~Roulet,
\newblock Phys.Lett. {\bf B399}, 113 (1997), [hep-ph/9611425].

\bibitem{Pilaftsis:1997jf}
A.~Pilaftsis,
\newblock Phys.Rev. {\bf D56}, 5431 (1997), [hep-ph/9707235].

\bibitem{joshipura:1992hp}
A.~S. Joshipura and J.~W.~F. Valle,
\newblock Nucl. Phys. {\bf B397}, 105 (1993).

\bibitem{Aad:2014iia}
ATLAS Collaboration, G.~Aad {\em et~al.},
\newblock 1402.3244.

\bibitem{Romao:1991tp}
J.~C. Romao, N.~Rius and J.~W.~F. Valle,
\newblock Nucl. Phys. {\bf B363}, 369 (1991).

\bibitem{Hirsch:2009ee}
M.~Hirsch, A.~Vicente, J.~Meyer and W.~Porod,
\newblock Phys.Rev. {\bf D79}, 055023 (2009), [0902.0525].

\bibitem{GarciaiTormo:2011et}
X.~Garcia~i Tormo, D.~Bryman, A.~Czarnecki and M.~Dowling,
\newblock Phys.Rev. {\bf D84}, 113010 (2011), [1110.2874].

\end{thebibliography}

\end{document}